\documentstyle[preprint,prl,aps]{revtex}
\def\degres{\mbox{$^\circ$}}      % mathematical symbols

\def\gsim{\mbox{$\stackrel{_>}{_\sim}$}} %Maggiore o circa%
\def\bec{\begin{center}}
\def\eec{\end{center}}

\def\beq{\begin{equation}}
\def\eeq{\end{equation}}

\renewcommand{\frac}[2]{{{\displaystyle #1}\over{\displaystyle #2}}}

\def\calS{\mbox{$\cal S$}}

\def\hath{\mbox{${\hat{h}}$}}

\def\dms{\mbox{$\Delta m^2$}}     % neutrino MSW parameters

\def\SdTvS{\mbox{$\sin^2 2\theta_V$}}

\def\numt{\mbox{$\nu_{\mu(\tau)}$}} %  nu_mu(tau) neutrinos
\def\nue{\mbox{$\nu_e$}}            %  nu_e
\def\nus{\mbox{$\nu_s$}}            %  Sterile neutrino
\def\TYear{\mbox{$T_{y}$}}

\def\YeCore{\mbox{$Y_e(core)$}}       %    "       "   in the Core
\def\TeTh{\mbox{$T_{e,th}$}}     % Threshold Electron energy
\def\AsymRs{\mbox{$A_{D-N}^s$}}     %        "      with s index
\def\AsymRN{\mbox{$A_{D-N}^N$}}     %      night
\def\AsymRC{\mbox{$A_{D-N}^C$}}     %      core
\def\Ps{\mbox{${\bar{P}}_\odot$}}       % Survival probability in the Sun
\def\PeTw{\mbox{$P_{e2}$}}              % Transition nu_2 -> nu_e
\def\TResidN{\mbox{$T_{res}^N$}}      %   Night
\def\TResidC{\mbox{$T_{res}^C$}}      %   Core
\def\TResidM{\mbox{$T_{res}^M$}}      %   Mantle
\def\night{\mbox{\em{Night}}}                    %    Night
\def\core{\mbox{\em{Core}}}                      %    Core
\def\mantle{\mbox{\em{Mantle}}}                  %    Mantle
\def\SK{Super - Kamiokande}
\def\FORTRAN{\mbox{\tt{FORTRAN}}}

\def\deg{\degres}  % degrees symbol

\def\ltap{\ \raisebox{-.4ex}{\rlap{$\sim$}} \raisebox{.4ex}{$<$}\ }
\def\gtap{\ \raisebox{-.4ex}{\rlap{$\sim$}} \raisebox{.4ex}{$>$}\ }

\def\ltap{\lsim}  % Less or similar to

\def\SNOLat{\mbox{$\lambda_{\mathrm{SNO}}$}}

\def\ltap{\ \raisebox{-.4ex}{\rlap{$\sim$}} \raisebox{.4ex}{$<$}\ }
\def\gtap{\ \raisebox{-.4ex}{\rlap{$\sim$}} \raisebox{.4ex}{$>$}\ }

\begin{document}
\sloppy

\vspace{0.5cm}
%
% References Pointer
%
%{\normalsize
% \begin{flushright}
% \begin{tabular}{l}
% SISSA ? /99/EP\\
% hep-ph/9908xxx\\
% August, 1999\\
% \vspace{1cm}
%%%%%%%%%%%%%
% The Title %
%%%%%%%%%%%%%
\draft
\begin{titlepage}
\newpage
\preprint{\vbox{\baselineskip 10pt{
\hbox{Ref. SISSA 30/2000/EP}
\hbox{March, 2000}
% \hbox{hep -- ph/000xxx}
}}}
\vskip -0.4cm
\title{\bf Day - Night Effect Predictions for the SNO Detector}
\author{M. Maris$^{a}$ and S.T. Petcov $^{b,c)}$\footnote{Also at:
Institute of Nuclear Research and Nuclear Energy, Bulgarian Academy
of Sciences, BG--1784 Sofia, Bulgaria.}}
\address{a) Osservatorio Astronomico di Trieste, I-34113 Trieste, Italy\\
\vskip -0.4cm
% \address{
b) Scuola Internazionale Superiore di Studi Avanzati, I-34014 Trieste, Italy\\
% }
\vskip -0.4cm
% \address{
c) Istituto Nazionale di Fizica Nucleare, Sezione di Trieste, I-34014
Trieste, Italy}
\vskip -0.4cm
\maketitle
\begin{abstract}
% \begin{minipage}{5in}
\baselineskip 16pt
\tightenlines 
 Detailed predictions for the day-night (D-N) asymmetry 
in the energy-integrated one year signals in the SNO detector
in the case of the MSW
$\nu_e \rightarrow \nu_{\mu(\tau)}$
and/or $\nu_e \rightarrow \nu_{s}$ transition solutions
of the solar neutrino problem are presented. 
The asymmetries in the
charged current (CC) and $\nu - e^{-}$ elastic scattering (ES)
event rates are calculated for both MSW solutions;
in the case of the $\nu_e \rightarrow \nu_{s}$ transition solution
the D-N asymmetry in the neutral current (NC) event rate 
are derived as well.
The asymmetries are calculated for 
three night samples of events which are produced by
the solar neutrinos crossing i) the Earth
mantle only ({\it Mantle}), ii) the Earth core ({\it Core}) 
and iii) the Earth core and/or the mantle ({\it Night}).
The effects of the uncertainties
i) in the values of the cross-sections of the CC and NC
neutrino-induced reactions on deuterium, and 
ii) in the value of the bulk matter
density and/or the chemical composition of the Earth core,
on the corresponding D-N asymmetry predictions 
are analyzed. It is shown, in particular, 
that that due to the strong enhancement of 
the transitions of the solar 
neutrinos crossing the Earth core, at
$\sin^22\theta_V \leq 0.01$
the corresponding one year average 
D-N asymmetry in the {\it Core} sample
of CC events in the case of the 
$\nu_e \rightarrow \nu_{\mu(\tau)}$
solution can be larger 
by a factor of up to $\sim 8$ 
than the asymmetry 
in the {\it Night} sample.
In certain subregions of the MSW
solution regions at small 
$\sin^22\theta_V$,
the predicted magnitude
of the {\it Core} D-N asymmetry
in the CC sample is very sensitive
to the value of the electron fraction number 
in the Earth core.
Iso - (D-N) asymmetry contours in the $\dms - \SdTvS$
plane for the SNO detector are derived in the
region $\SdTvS \gsim 10^{-4}$ for 
the {\it Core} and {\it Night} 
% and {\it mantle} 
samples of the CC, ES and NC events.
The dependence of the D-N asymmetries
considered
on the final state e$^{-}$ threshold energy in the
CC and ES reactions is also investigated.
Our results show, in particular,
that the SNO experiment will be able
to probe substantial parts of 
the SMA and LMA MSW 
$\nu_e \rightarrow \nu_{\mu(\tau)}$
solution regions be performing
{\it Night} and {\it Core}
D-N asymmetry measurements.
% \end{minipage}
\end{abstract}
\vspace{3mm}
% \indent
% {\underline{PACS:} 14.60Pq, 26.65, 95.85.Ry}

\end{titlepage}
\newpage
\tightenlines
\leftline{\bf I. Introduction}

    In spite of the remarkable progress 
made in the studies of solar neutrinos 
% during the last nine years or so,
the true cause of the solar neutrino deficit observed
in the $Cl - Ar$, Kamiokande, $Ga - Ge$,  and 
Super-Kamiokande experiments \cite{Cl,K,SAGE,GALLEX,SK99} 
is still not identified. The existing helioseismological data
and its interpretation \cite{Helios}
make very unlikely the possibility of 
an astrophysical origin
of the discrepancy between the solar 
neutrino observations \cite{Cl,K,SAGE,GALLEX,SK99} and 
the standard solar model predictions \cite{JNB,BP98}.
The hypotheses of vacuum oscillations
or MSW transitions of solar neutrinos continue 
to provide viable solutions of the 
solar neutrino problem 
(see, e.g., \cite{SK99,Fogli99,CGG99,PK99}). 

  The current (mean event rate) 
solar neutrino data admit three types of 
MSW solutions  
if the solar $\nu_e$ undergo two-neutrino 
transitions into active neutrinos,
$\nu_e \rightarrow \nu_{\mu (\tau)}$:
the well-known small mixing angle 
(SMA) non-adiabatic
and large mixing angle (LMA) adiabatic 
(see, e.g., \cite{KPMSW93,HataL94}) and 
the so-called ``LOW'' solution 
(very recent analyses can be found 
in, e.g., \cite{SK99,Fogli99,CGG99,PK99}).
While the SMA and LMA solutions have been shown 
to be rather stable with respect to variations 
in the values of the various physical quantities which 
enter into the calculations (the fluxes of 
$^8$B and $^7$Be neutrinos, 
nuclear reaction cross-sections, etc.),
and of the data utilized in the analyses,
the LOW solution is of the ``borderline'' 
type: its existence even at 99\% C.L. 
is not stable with respect to
relatively small changes in the data 
and/or in the 
relevant theoretical predictions 
(see, e.g., \cite{Fogli99} and 
the references quoted therein.).
To the three solutions there correspond
three distinct regions in the plane of values 
of the two parameters, 
$\Delta m^2$ and $\sin^22\theta$, 
characterizing the transitions.
One finds \cite{SK99,Fogli99,PK99} using the 
standard solar model predictions 
\cite{BP98}  for the solar neutrino fluxes 
($^8$B, $^7$Be, $pp$, etc.) 
that at 99\% C.L. the SMA MSW solution requires
values in the intervals 
$4.0 \times 10^{-6}~{\rm eV^2} \ltap \Delta m^2 \ltap  
10.0\times 10^{-6}~{\rm eV^2}$,
$1.3 \times 10^{-3} \ltap \sin^22\theta_V 
\ltap 1.0\times 10^{-2}$,
the LMA solutions is realized for
$\Delta m^2$ and $\sin^22\theta_V$
from the region
$7.0\times 10^{-6}~{\rm eV^2} \ltap \Delta m^2 \ltap  
2.0\times 10^{-4}~{\rm eV^2}$,
$0.50 \ltap \sin^22\theta_V 
\ltap 1.0$, and the LOW solution 
lies approximately in the region
$0.4\times 10^{-7}~{\rm eV^2} \ltap \Delta m^2 \ltap  
1.5\times 10^{-7}~{\rm eV^2}$,
$0.80 \ltap \sin^22\theta_V 
\ltap 1.0$. 
% % \bec\beq\label{eq:kpnu96:a}
% %  \begin{array}{c}
% $$  3.6 \times 10^{-6}~{\rm eV^2} \ltap \Delta m^2 \ltap  
% 9.8\times 10^{-6}~{\rm eV^2},~~\eqno(1a)$$
% $$  4.5 \times 10^{-3} \ltap \\sin^22\theta 
% \ltap 1.3\times 10^{-2},~~\eqno(1b)$$
% %  \end{array}
% % \eeq\eec
% \noindent or
%% \bec\beq\label{eq:kpnu96:b}
%%  \begin{array}{cc}
%    5.7 \times 10^{-6} \mbox{ eV}^2 \, \ltap \, \Delta m^2 \,
%                                     \ltap\, 9.5\times 10^{-5} \mbox{ eV}^2,\\
%    \\
%    0.51 \,\ltap \, \SdTvS \, \ltap\, 0.92, \\
%%  \end{array} 
%% \eeq\eec
% \noindent
The SMA and LMA solution regions expand
in the direction of smaller values of 
$\sin^22\theta_V$ up to
 $\sim 0.6\times 10^{-3}$ and 
to $\sim 0.3$, respectively, 
if one adopts more conservative approach 
in analyzing the data in terms of the MSW effect 
and treats the $^8$B neutrino flux as a free parameter
in the analysis (see, e.g., \cite{PK99}). 
This enlargement of the two solution regions 
is not uniform in $\Delta m^2$ (see further).
We will use the term ``conservative''
for the MSW solution regions derived
by treating the  $^8$B neutrino 
flux as a free parameter. 

   The solar neutrino data can also be explained assuming
that the solar neutrinos undergo MSW transitions into
sterile neutrino in the Sun: $\nu_e \rightarrow \nu_s$ (see, e.g,
\cite{PK99,KLPSter96}). In this case only a SMA solution
is compatible with the data. The corresponding 
solution region obtained (at a given C.L.) 
using the mean event rate solar neutrino data 
and the predictions of
ref. \cite{BP98} for the different 
% components ($^8$B, $^7$Be, $pp$, $pep$, $CNO$) of 
solar neutrino flux components
practically 
coincides in magnitude and shape 
with the SMA $\nu_e \rightarrow \nu_{\mu (\tau)}$
solution region, but is shifted 
by a factor of $\sim 1.2$
along the $\Delta m^2$ axis to smaller 
values of $\Delta m^2$.
The ``conservative'' $\nu_e \rightarrow \nu_s$ 
transition solution region extends both 
in the direction of smaller and larger 
values of $\sin^22\theta_V$ down to
$0.7\times 10^{-3}$ and up to 0.4~ \cite{PK99}.

  The results on the spectrum of the final state $e^{-}$ 
from the $\nu - e^{-}$ elastic scattering reaction,
obtained in the range of the  recoil$-e^{-}$ energy
$E_e \sim (5 - 20)~$MeV
by the Super-Kamiokande collaboration 
do not allow to rule out definitely 
any of the solutions
of the solar neutrino problem indicated
above \cite{SK99}. 
The observed rise of the spectrum at 
$E_e \gtap 12~$MeV can be due
to a contribution from the $hep$ solar neutrinos 
% produced in the reaction
% $p + ^{3}He \rightarrow ^{4}He + e^{+} + \nu_e$,
\cite{BK98},
which complicates the interpretation of the data.
The spectrum measured 
below 12 MeV is compatible with 
an energy-independent
suppression of the $^8$B neutrino flux,
or with a mild energy dependence 
of the suppression.

  A unique testable prediction of the 
MSW solutions of the solar 
neutrino problem is the day-night (D-N) effect -
% effects of the
% Earth matter on the neutrino transitions.
a difference between the 
solar neutrino event rates
% observed 
during the day and during the night, 
caused by the additional transitions of the solar
neutrinos taking place at night while the neutrinos 
cross the Earth on the way to the detector 
(see, e.g., \cite{HataL94,DNold,GelbKR97} 
and the references quoted therein).
The experimental observation of a non-zero  D-N asymmetry
\begin{equation}
A^{N}_{D-N} \equiv \frac{R_{N} - R_{D}}{(R_N + R_D)/2},
%~\eqno(1)$$
\end{equation}
\noindent where $R_N$ and $R_D$ are, e.g., the  
one year averaged event rates in a given detector 
respectively during the night and the day, 
would be a very strong evidence in favor 
(if not a proof) of the MSW solution
of the solar neutrino problem (see further). 

  Extensive predictions 
for the magnitude of the D-N effect
for the Super-Kamiokande detector have been obtained 
in \cite{SK97I,SK97II,SK98III,LisiM97,BK97}.
High precision calculations of
% the D-N effect were performed. Extensive predictions for 
the D-N asymmetry in the 
one year averaged recoil-e$^{-}$ spectrum and  
in the energy-integrated event rates were performed
for three event samples, 
{\it Night}, {\it Core} and {\it Mantle},
in \cite{SK97I,SK97II,SK98III}.
The night fractions of these event samples 
are due to neutrinos which respectively cross 
the Earth along any trajectory, 
cross the Earth core, and
cross only the Earth mantle (but not the core),
on the way to the detector. 
The measurement of the D-N asymmetry in the 
{\it Core} sample was found to be
of particular importance \cite{SK97I,SK97II} 
because of the 
strong enhancement of the asymmetry,
caused by a constructive interference  
between the amplitudes of the
neutrino transitions in the Earth 
mantle and in the Earth core 
\cite{SPPLB43498,ChPet991,ChPet992}. 
% The {\it mantle-core enhancement} of the D-N effect
% was quantitatively established and
% studied in great detail in \cite{SK97I,SK97II}.
The effect differs from the MSW one \cite{SPPLB43498}.
The  {\it mantle-core enhancement effect} 
is caused by the existence (for a given neutrino trajectory
through the Earth core) of 
{\it points of resonance-like 
total neutrino conversion}
in the corresponding space 
of neutrino oscillation 
parameters \cite{ChPet991,ChPet992}. 
The location of these points determines the regions
where the relevant probability of transitions
in the Earth of the Earth-core-crossing solar neutrinos
is large \footnote{Being a
constructive interference effect between 
the amplitudes of neutrino transitions 
in the mantle and in the core,
this is not just ``core enhancement'' effect,
but rather {\it mantle-core enhancement} effect.}
\cite{ChPet992}.
At small mixing angles 
and in the case of 
$\nu_e \rightarrow \nu_{\mu (\tau)}$ 
transitions
the predicted D-N asymmetry in the 
{\it Core} sample of the Super-Kamiokande
data was shown \cite{SK97II} to 
be much bigger due to the 
{\it mantle-core enhancement effect}
\footnote{The term ``neutrino oscillation length
resonance'' (NOLR) was used in \cite{SPPLB43498},
in particular, to denote the mantle-core enhancement 
effect in this case.}
-  by a factor of up to $\sim 6$,
than the asymmetry in the {\it Night} sample.
The asymmetry in the {\it Mantle} 
sample was found to be smaller than the 
asymmetry in the {\it Night} sample. 
On the basis of these results it was 
concluded in \cite{SK97II} that 
it can be possible to test a substantial part of the
MSW $\nu_e \rightarrow \nu_{\mu (\tau)}$ SMA 
solution region in the $\Delta m^2 - \sin^22\theta_V$
plane by  performing selective, i.e., 
{\it Core} D-N asymmetry measurements.

  The current Super-Kamiokande data \cite{SK99} shows a    
D-N asymmetry in the {\it Night} sample, which is 
different from zero at 1.9 s.d. level: 
\begin{equation}
A^{N}_{D-N} = 0.065 \pm 0.031~(stat.) \pm 0.013~(syst.).
% ~~\eqno(2)$$
\end{equation}
\noindent These data do not allow to test the SMA solution region: 
the predicted asymmetry is too small 
(see, e.g., \cite{SK97II,LisiM97,BK97}).
However, the Super-Kamiokande night data is given in
5 bins and 80\% of the events in the bin N5  
are due to Earth-core-crossing solar neutrinos \cite{SK99},
while the remaining 20\% are produced by neutrinos which cross 
only the Earth mantle. Since the predicted D-N asymmetry
in the {\it Mantle} sample is practically negligible 
in the case of the MSW SMA solution of interest \cite{SK97II},
we have for the D-N asymmetry measured using the  
night N5 bin data: $A^{N5}_{D-N} \cong 0.8A^{C}_{D-N}$,
$A^{C}_{D-N}$ being the asymmetry in the 
{\it Core} sample. The data on $A^{N5}_{D-N}$ \cite{SK99}
permitted to exclude a part of the MSW SMA solution region 
located in the area
$\sin^22\theta_V \cong (0.007 - 0.01)$,
$\Delta m^2 \cong (0.5 - 1.0)\times 10^{-5}~{\rm eV^2}$.
% which is allowed by the mean event rate solar neutrino data. 
Let us note that it was possible to probe 
the indicated MSW SMA solution region
only due to the {\it mantle-core enhancement} 
\cite{SPPLB43498,ChPet991,ChPet992}
of the the {\it Core} part of the asymmetry 
$A^{N5}_{D-N}$.
It should be obvious from the above discussion that
the measurement of the {\it Core} asymmetry
$A^{C}_{D-N}$, as suggested in \cite{SK97II},
will provide a more effective  test
of the the MSW SMA solution
\footnote{We were happy to learn \cite{YSSK1299}  
that recently the Super-Kamiokande collaboration has decided 
to present experimental results on the {\it Core} 
D-N asymmetry as well.}
than the measurement of $A^{N5}_{D-N}$.

 As was shown in \cite{SK98III,SPPLB43498},
the {\it mantle-core enhancement} is not so effective 
in the $^{8}$B neutrino energy interval 
$E \sim (5 - 12)~$MeV when the solar neutrino
transitions are of the type 
$\nu_e \rightarrow \nu_s$: 
it does not produce such a dramatic enhancement 
of the {\it Core} D-N asymmetry 
as in the case of $\nu_e \rightarrow \nu_{\mu (\tau)}$ 
transitions. More generally, the
measurements of the D-N effect related observables 
in the Super-Kamiokande experiment appears 
at present unlikely 
to provide an effective test of 
the MSW $\nu_e \rightarrow \nu_s$ (SMA)
solution of the solar neutrino problem \cite{SK98III}.

   The studies of the D-N effect will be continued by the SNO experiment
which is already operational \cite{SNO}. As is well known,
the solar neutrinos can be detected in SNO via the charged current (CC)
reaction on deuterium,
\begin{equation}
\nu_e + D \rightarrow e^{-} + p + p,~~~~~~~(CC)
\end{equation}
\noindent deuterium break up by neutrinos via the neutral current (NC)
weak interaction,
\begin{equation}
\nu + D \rightarrow \nu + n + p,~~~~~~~~(NC) 
% ~~\eqno(4)$$
\end{equation}
\noindent and via the elastic scattering (ES) on electrons, 
\begin{equation}
\nu + e^{-} \rightarrow \nu + e^{-}.~~~~~~~~(ES)
% ~~\eqno(5)$$ 
\end{equation}
   In the present article
we derive detailed predictions for the D-N asymmetries in 
the CC, NC and ES event rates for the SNO detector. 
For the $^8$B neutrino flux, $\Phi(B)$, 
which at the Earth surface 
represents 47.5\% of the flux predicted by the 
standard solar model \cite{BP98}, $\Phi_{SSM}(B)$,
the expected number of events per year 
due to the CC reaction
$\nu_e + D \rightarrow e^{-} + p + p$
in the SNO detector
is about \cite{SNO} 3300
% (or 12.7 events per day)
for $T_{e,th} = 5$ MeV, $T_{e,th}$ being the threshold
kinetic energy of the final state $e^{-}$. 
This is comparable to the
event rate in the \SK\ experiment. 
The expected event rate \cite{SNO} 
due to the NC reaction
$\nu_l + D \rightarrow \nu_l + n + p$
for a full standard solar model \cite{BP98}
$^8$B neutrino flux (and 45\% detection efficiency)  
is approximately 
(5.5 - 6.0) events per day,
while the event rate due to the $\nu - e^{-}$
elastic scattering (ES) reaction is predicted 
to be $\sim 1.4$ events per day
for $\Phi(B) \sim 0.47 \Phi_{SSM}(B)$
and $T_{e,th} = 5$ MeV.
The indicated statistics 
permits to perform a high precision
search for the D-N effect 
and for the Earth {\it mantle-core enhancement}
of the effect
in the sample of events due to the CC reaction.
Therefore the main emphasis of our study will be
on the predictions for the magnitude of the D-N 
effect in the CC sample of events in the SNO detector,
although rather detailed results for the 
expected magnitude of 
the effect in the ES and NC samples will 
also be presented. Let us remind the reader that
the D-N asymmetry in the NC event rate can be nonzero
in the case of the MSW solution of the solar neutrino problem
with transitions into sterile neutrino \cite{KLPSter96}.

  Predictions for the D-N effect
% related observables
for the SNO detector were derived  in 
refs. \cite{HataL94,DNold,LisiM97,BK97} and recently in \cite{BKS00}.
However, our study overlaps little with 
those performed in \cite{HataL94,DNold,LisiM97,BK97,BKS00}.

\vskip 0.3cm
\leftline{\bf 2. Calculating the D-N Effect for the SNO Detector}
\vskip 0.3cm
\leftline{\bf 2.1. The CC and NC Reaction Cross Section Uncertainties}
\vskip 0.3cm

   The high precision calculations 
of the D-N effect for the SNO detector performed in the 
present article
are based on the methods developed for our
studies of the D-N effect for the Super-Kamiokande detector, 
which are described in detail in \cite{SK97I,SK97II,SK98III}.
The cross sections of the CC and NC reactions on deuterium
were  taken from \cite{XS:IAUTH0,XS:IAUTH1,XS:IAUTH2}. They were 
tabulated using the 1996 version of the E. Lisi 
$\nu\mathrm{d}$ \FORTRAN\
library \cite{Cross:Sources:For}. 
The cross section of the
$\nu_e - e^-$ elastic scattering reaction 
was taken from \cite{ESCross}.
We used in our calculations 
the $^8$B neutrino spectrum  
derived in 
\cite{B(8)nuSpectrum}. 
The probability
of survival of the solar $\nu_e$ when they travel 
in the Sun and further to the surface of the Earth,
$\bar{P}_{\odot}(\nu_e \rightarrow \nu_e)$,
was computed on the basis of the 
analytic expression obtained in \cite{SPMSWExp}
and using the method developed in
\cite{KPMSW88}. In the calculation of 
$\bar{P}_{\odot}(\nu_e \rightarrow \nu_e)$
we have utilized (as in \cite{SK97I,SK97II,SK98III})  
the density profile 
of the Sun and the $^8$B neutrino production 
distribution in the Sun, predicted in \cite{BP95}.
These predictions were updated in \cite{BP98},
but a detailed test study showed 
\cite{Maris:IRP:1999} (see also \cite{BK97})
that using the results of \cite{BP98} instead 
of those in \cite{BP95} leads to a change of 
the survival probability by less than 1\%
for any set of values of the parameters
$\Delta m^2$ and $\sin^22\theta$, 
relevant for the calculation of the D-N effect. 
As in \cite{SK97I,SK97II,SK98III},
the Earth matter effects were calculated 
using the Stacey model from 1977 \cite{Stacey77}
for the Earth density distribution. 
The latter practically coincides with 
that predicted by the more recent Earth model 
\cite{PREM81}. Most of the results
are obtained for two sets of values of
the electron fraction number in the Earth mantle and
the Earth core, 
$Y_e(man)$ and $Y_e(core) \equiv Y_{e}^{c}$, which reflect
the chemical composition of 
the two major Earth structures:
the standard ones $Y_e(man) = 0.49$,
$Y_e^c = 0.467$, and for the possibly 
conservative upper bound of  $Y_e^c = 0.500$ 
(see, e.g, \cite{YeEarth,SK97II}). The latter effectively
accounts for the uncertainties both in $Y_e^c$ 
and in the average matter density of the Earth core.  

  Three different  calculations of the cross section of
the $\nu_e - D$ CC reaction are available in the literature
\cite{XS:IAUTH0,XS:IAUTH1,XS:IAUTH2}.
In two of these articles 
\cite{XS:IAUTH1,XS:IAUTH2} the cross section of the 
$\nu_l - D$ NC reaction was also computed.
Extensive numerical study has shown
that i) the {\it maximal} differences between the various 
one year averaged 
D-N asymmetries in the CC event rate, 
calculated using the three different predictions 
for the CC cross section, 
% lie in the range of $1\% \div 6\%$,  
do not exceed 6\%,
ii) the differences in the range of (1 - 6)\%
occur only in a set of disconnected regions in the
$\Delta m^2 - \sin^22\theta_{V}$ plane,
which are essentially point-like and 
whose total area is exceedingly small, and
iii) these small point-like regions of maximal spread  
in the D-N asymmetry predictions are generally located 
outside and sufficiently far from 
the MSW solution regions of interest 
(including the ``conservative'' ones). 
There is only one case when such a difference 
occurs on the borderline of 
an MSW  ``conservative'' solution region:
we found a spread in the prediction
for D-N asymmetry in the \core\ sample of events
of $\sim 2.5\%$ in the case of
$\nu_e \rightarrow \numt$ transitions 
for $\sin^22\theta\sim 0.6$ and 
$\Delta m^2 \sim 3 \times 10^{-5}$ eV$^2$,
which is at the border of the LMA 
``conservative'' solution region.

     The cross sections 
of the $\nu_e - D$ CC reaction calculated in
\cite{XS:IAUTH0,XS:IAUTH1,XS:IAUTH2}
differ by $\sim$ (5 - 15)\% 
at the  $^8$B neutrino energies of interest. 
The question of why the predicted 
D-N asymmetries in the CC and NC event rates 
are so insensitive to the choice of the 
cross section used in the calculations naturally
arises. The product of the 
normalized to one
$^8$B neutrino spectrum,
$\calS(E)$, and of the CC reaction cross section,
$\sigma_{CC}(E)$, which enters into 
the expression for the CC event rate,
can be approximated by a resonance-like  
function defined by its full width
at half maximum, the position of the maximum
in $E$ and by its value at the maximum.
These parameters were calculated for the 
three different cross sections 
\cite{XS:IAUTH0,XS:IAUTH1,XS:IAUTH2} and
for different values of the threshold 
kinetic energy of the final state
electron, $T_{e,th}$. The results are 
given in Table I. 
% \ref{tab:CSTAB}.
The largest 
differences between the three results for 
$\calS(E)\sigma_{CC}(E)$
% \cite{XS:IAUTH0,XS:IAUTH1,XS:IAUTH2} 
happen in the fastly decreasing part of 
the $^8$B neutrino spectrum ($E \gtrsim 10~$ MeV). 
As Table I shows, the spread in the values at the maximum 
is between 3.5\%
and 10\%, while the position of the maximum 
in $E$ and the F.W.H.M. change at most by $0.8\%$.
Consequently, the differences between the
three CC cross sections 
\cite{XS:IAUTH0,XS:IAUTH1,XS:IAUTH2} 
lead essentially only to a non-negligible
change in the prediction for 
the overall normalization factor for the
total CC event rate, which practically 
does not affect the D-N asymmetry. 
The same conclusions 
can be shown to be valid for the
implications of the differences 
between the two NC cross sections 
\cite{XS:IAUTH1,XS:IAUTH2}
for the uncertainties 
in the predicted values of the
D-N asymmetries 
in the NC signal in SNO in the case of 
MSW $\nu_e \rightarrow \nu_s$ transition
solution of the solar neutrino problem.
These results are illustrated in Figs. 1a - 1f,
where the difference between the values of the 
{\it Core} ({\it Night}) D-N asymmetry
calculated utilizing the predictions for the 
relevant CC and NC reaction cross-sections
derived in ref. \cite{XS:IAUTH1} and in 
ref. \cite{XS:IAUTH2} are shown as functions 
of $\Delta m^2$ and $\sin^22\theta_V$ 
in the intervals ($10^{-7} - 10^{-4}$) eV$^2$
and ($10^{-4} - 1.0$), respectively
\footnote{However, all the
plots produced in the framework of this study 
are at disposition of the interested 
reader at the site {\tt
http://www.pv.infn.it/$\sim$maris/nue/sno/index.html}.}.

This analysis permits one to conclude 
that even neglecting
the electron energy resolution function and the
expected statistical and systematical 
errors in the measurements, 
the calculation of the D-N asymmetries in the 
CC and NC event rates in the SNO detector 
utilizing the different 
cross sections from
\cite{XS:IAUTH0,XS:IAUTH1,XS:IAUTH2}
should give the same results within 1\% 
in the MSW ``conservative'' solution regions.
Accordingly, we will
present predictions for the D-N asymmetries 
in the CC and NC signals computed
using the cross sections 
of ref. \cite{XS:IAUTH1}.

\vskip 0.3cm
\leftline{\bf 2.2 Comments on the Probability of Solar $\nu_e$ Survival}
\vskip 0.3cm

  The SNO detector is located at $\SNOLat = 46\deg \, 20'$\ North,
$\lambda'_{SNO} = 81\deg \, 12'$\ West. At the indicated
latitude we have:
$\TResidC/\TYear = 0.0396$, where $T^{C}_{res}$ is the 
{\it Core} residence time, i.e., the total time in one year
during which the solar neutrinos cross the Earth core at night 
on the way to the 
SNO detector, $T_{y} = 365.24$ days.
This implies that at the location 
of the SNO experiment
only for $7.9\%$ of the
one year night time the Sun is behind 
the Earth core with respect to SNO, to be
compared with  $14.85\%$ for the  \SK\ detector
(see, e.g., \cite{SK97I}). Thus, the 
statistical error in the measurement of the 
{\it Core} D-N asymmetry with SNO detector 
will be approximately by a 
factor of 3.55 bigger than in the measurement  
of the {\it Night} D-N asymmetry.
In the case of the \SK\ detector this factor is 2.66.
Let us note also that for the {\it Mantle} residence time,
$\TResidM$, i.e., the total time in one year
during which the solar neutrinos cross at 
night the Earth mantle
but do not cross the Earth core  
on the way to the SNO detector, we have   
$\TResidM/\TYear = 0.4619$, and that the {\it Night} residence
time is $\TResidN = 0.5015\TYear$. 

  As is well-known,  
one can use the $\nu_2 \rightarrow \nu_e$ transition 
probability, $P_{e2}$, to account for  
the Earth matter effects 
in the solar $\nu_e$ survival 
probability,
$\bar{P}(\nu_e \rightarrow \nu_e)$,
in the case of the MSW solutions 
of the solar neutrino problem.
During the day one has 
$\bar{P}(\nu_e \rightarrow \nu_e) = 
\bar{P}_{\odot}(\nu_e \rightarrow \nu_e)$,
where $\bar{P}_{\odot}(\nu_e \rightarrow \nu_e)$
was defined earlier as the probability of 
survival of the solar $\nu_e$
when they travel to the Earth surface 
without traversing the Earth.
A nonzero difference between 
$\bar{P}(\nu_e \rightarrow \nu_e)$
for the Earth-crossing neutrinos
and $\bar{P}_{\odot}(\nu_e \rightarrow \nu_e)$
produces a nonzero D-N asymmetry.
It is possible to have 
$\bar{P}(\nu_e \rightarrow \nu_e)
- \bar{P}_{\odot}(\nu_e \rightarrow \nu_e) \neq 0$
when i) $\bar{P}_{\odot}(\nu_e \rightarrow \nu_e) \neq 0.5$
and ii) $P_{e2}$ differs from its value in vacuum,
$P_{e2} \neq \sin^2\theta_V$ (for further details see, e.g., 
\cite{SK97I,SK97II,SK98III,LisiM97,BK97}).

   We have calculated the one year averaged 
$\nu_2 \rightarrow \nu_e$ transition 
probability for the location of the SNO detector, $<P_{e2}>$, 
using the methods described in 
\cite{SK97I}. As in \cite{SK97I},
this was done for each of the three different event samples
{\it Core}, {\it Night} and {\it Mantle} we are considering,
$<P_{e2}>^{C,N,M}$. We comment below only on few specific
features of the probabilities $<P_{e2}>^{C,N,M}$
and of the corresponding solar $\nu_e$ survival
probabilities at night, $\bar{P}^{C,N,M}(\nu_e \rightarrow \nu_e)$.

\vskip 0.3cm
\leftline{\bf 2.2.1 $\nue \rightarrow \numt$\ Transitions}
\vskip 0.3cm

  At small mixing angles, $\sin^22\theta_V \ltap 0.01$, 
$<P_{e2}>^{C}$ is considerably larger 
than $<P_{e2}>^{N,M}$ in most of the interval of 
values of $E_\nu/\Delta m^2$ of interest: the absolute
maximum of $<P_{e2}>^{C}$, for instance, exceeds
the absolute maxima of $<P_{e2}>^{N,M}$ by
a factor of $\sim (4 - 5)$. 
This is illustrated in Figs. 2a - 2b.  
The strong resonance-like enhancement of  
$<P_{e2}>^{C}$ with respect to
$<P_{e2}>^{N,M}$,
seen, in particular, in Figs. 2a - 2b, is 
\footnote{The plots of $<P_{e2}>^{C,N,M}$ as functions of
$\Delta m^2/E$ at fixed $\sin^22\theta_V \ltap 0.01$
are very similar to those for the Super-Kamiokande
detector, published in ref. \cite{SK97I} which
includes a large set of such plots.}
due to a constructive interference
between the amplitudes of the 
neutrino transitions in the Earth mantle and 
in the core \cite{SPPLB43498,ChPet991,ChPet992}.
The interference produces 
points of total neutrino conversion
\cite{ChPet991,ChPet992}, 
$\mathrm{max}(P_{e2}) = 1$.
For each fixed $\hat{h}$ 
from the interval $\hat{h} = (0^{\circ} - 30^{\circ})$,
there exists one such point 
in the region $\sin^22\theta_V \leq 0.10$,
located in the interval
$\sin^22\theta_V \cong (0.04 - 0.08)$ 
(see Table 4 in \cite{ChPet992}).
At $\hat{h} = 23^{\circ}$, for instance, 
we have $\mathrm{max}(P_{e2}) = 1$
at $\sin^22\theta_V = 0.06$ 
and $\Delta m^2/E = 6.5\times 10^{-7}~{\rm eV^2/MeV}$~
\footnote{ 
The points under discussion
change somewhat their position 
with the change of $\hat{h}$ 
(see Table 4 in \cite{ChPet992}).}.
The probability $P_{e2}$
is enhanced at each given 
$\hat{h} \lesssim 30^{\circ}$
in a sufficiently wide 
region in the neighborhood 
of the corresponding point
of total neutrino conversion, which causes 
the strong enhancement of $<P_{e2}>^{C}$ 
under discussion at
$\sin^22\theta_V \ltap 0.01$.

   For $\sin^22\theta_V \ltap 0.004$, 
$<P_{e2}>^{C,N,M}$ are nonnegligible 
in an interval of values of
$E/\Delta m^2$ where 
$\bar{P}_{\odot}(\nu_e \rightarrow \nu_e) > 0.5$
and the D-N asymmetries (defined as in eq. (1))
for all the three samples of events
{\it Core}, {\it Night} and {\it Mantle},
$A^{C,N,M}_{D-N}$, are negative. 
For the values of
$\Delta m^2$ from the SMA solution region
and the neutrino energies of interest,
$E \cong (5.0 - 14.4)~$MeV,
the core asymmetry $A^{C}_{D-N}$
goes through zero in the interval 
of $\sin^22\theta_V \cong (0.004 - 0.006)$.
For each given value of 
$\sin^22\theta_V$ from this interval
the zero D-N effect line
$\bar{P}_{\odot}(\nu_e \rightarrow \nu_e) = 0.5$
taking place at a specific value of
$E/\Delta m^2$, $(E/\Delta m^2)_{0}$,
splits $<P_{2e}>^{C}$  
into two parts:
the part at 
$E/\Delta m^2 > (E/\Delta m^2)_{0}$
generates a negative D-N asymmetry,
while the part at 
$E/\Delta m^2 < (E/\Delta m^2)_{0}$
creates a positive D-N asymmetry.
When integrating over the neutrino energy
the positive and the negative contributions
to $A^{C}_{D-N}$ tend to mutually
cancel. For each given
$\sin^22\theta_V \cong (0.004 - 0.006)$
there exists a $\Delta m^2$ 
from the SMA solution region
for which this cancellation is exact 
giving $A^{C}_{D-N} = 0$.
In the case of the {\it Night} 
asymmetry, the interval 
in $\sin^22\theta_V$
in which one has $A^{N}_{D-N} = 0$
for a certain value of 
$\Delta m^2$ from 
the SMA solution region is larger:
$\sin^22\theta_V \cong (0.004 - 0.008)$.
For $\sin^22\theta_V \gtap 0.006~(0.008)$
we have $A^{C}_{D-N} > 0$ 
($A^{N}_{D-N} > 0$).

 As in \cite{SK97I}, one can introduce 
the probability D-N asymmetries,
\begin{equation}
A_{P}^{C,N,M} = 
{{\bar{P}^{C,N,M}(\nu_e \rightarrow \nu_e) -
\bar{P}_{\odot}(\nu_e \rightarrow \nu_e)}\over{
0.5(\bar{P}^{C,N,M}(\nu_e \rightarrow \nu_e) +
\bar{P}_{\odot}(\nu_e \rightarrow \nu_e))}},
\end{equation}
which give an indication about the possible magnitude
of the corresponding energy-integrated
event rate asymmetries.  For
$\sin^22\theta_V < 0.004$, 
$A_{P}^{C,N}$ are essentially negative and 
we have $|A_{P}^{C}| \ltap 3.0\%;~4.7\%$,
while $|A_{P}^{N}| \ltap 1\%$.
At $\sin^22\theta_V = 0.004$,
$A_{P}^{C}$ and $A_{P}^{N}$  
as functions of $E/\Delta m^2$
have substantial positive and negative parts
and $ - 4\% \ltap A_{P}^{C} \ltap 5.0\%$,
$ - 1.5\% \ltap A_{P}^{N} \ltap 3.5\%$.
As $\sin^22\theta_V$ increases beyond the zero
asymmetry region, $A_{P}^{C}$ increases steadily while
the increase of $A_{P}^{N}$ is rather slow.
At $\sin^22\theta_V = 0.008;~0.010$, for instance,
$A_{P}^{C}$ reaches the values of $\sim 26\%;~50\%$ and 
$A_{P}^{N} \ltap 2.7\%;~8.0\%$. This corresponds to
an enhancement factors of the {\it Core} asymmetry
at the maxima of $\sim 9.6;~6.3$.
The above results suggests that for values of
$\sin^22\theta_V$ from the SMA solution region,
the {\it Core} event rate asymmetry
in SNO can be bigger than the {\it Night}
asymmetry by factors which
can exceed the similar factors 
for the predicted {\it Core} and {\it Night}
asymmetries in the Super-Kamiokande detector: 
in the latter case the {\it Core} asymmetry enhancement
factor does not exceed $\sim 6$ \cite{SK97I,SK97II}.

 It is interesting to note that 
% as already noted in \cite{SK97I},
for $\sin^22\theta_V \cong (4 - 10)\times 10^{-3}$,
$A_{P}^{C,N}$ have relatively large positive values 
at $E/\Delta m^2 \approx few~ \times 10^{-5}~{\rm MeV/eV^2}$.
In the same region  $A_{P}^{C,N}$ exhibit 
(as functions of $E/\Delta m^2$) fast 
small amplitude oscillations as well.
These maxima of $A_{P}^{C,N}$ are
responsible for the noticeable ``horn''
at $\sin^22\theta_V \gtrsim 5\times 10^{-3}$
and $\dms \gtrsim 2 \times 10^{-5}$\ eV$^2$,
shown by the CC and NC
D-N asymmetry contour plots in the case of the 
$\nue \rightarrow \numt$ 
and $\nu_e \rightarrow \nu_s$  transitions, 
as well as by the ES D-N 
asymmetry contour plots for 
$\nue \rightarrow \nus$\ transitions (see further).
The lack of a similar ``horn'' 
in the ES D-N asymmetry contour plots in the $\nue
\rightarrow \numt$\ case is due to the neutral current
contribution of $\nu_{\mu(\tau)}$ to the ES event rate. 
The indicated feature of  
$A_{P}^{C,N}$
affects the D-N asymmetries of interest only
for a set of neutrino parameters 
which lie outside the regions of the MSW solutions
and we are not going to discuss it further.

   The {\it mantle-core enhancement} of $A_{P}^{C}$ 
\cite{SPPLB43498,ChPet991,ChPet992}
% with respect to $A_{P}^{N,M}$ 
is much less dramatic in the region
of the LMA solution: actually, in this region 
we typically have $A_{P}^{C} \cong (1.0 - 1.25) A_{P}^{N}$,
with $A_{P}^{C}$ exhibiting as a function of
$E/\Delta m^2$ rather large amplitude 
oscillations which are absent
in $A_{P}^{N}$.

\vskip 0.3cm
\leftline{\bf 2.2.2 $\nue\rightarrow\nus$\ Transitions}
\vskip 0.3cm

    The Earth {\it mantle-core enhancement} of $P_{e2}$  
at small mixing angles 
% \cite{SPPLB43498, ChPet991,ChPet992},
is operative in the case of
$\nu_e \rightarrow \nu_s$ transitions
as well \cite{SK98III,SPPLB43498,ChPet991,ChPet992}. 
For $\sin^22\theta_V \cong (0.1 - 1.0)\times 10^{-2}$
the corresponding absolute maximum of 
the one-year averaged {\it Core} probability
$<P_{e2}>^{C}$ 
exceeds the absolute maximum of
$<P_{e2}>^{N}$ by a factor of $\sim 4$.
However, at 
$\sin^22\theta_V \lesssim 0.01$,
$<P_{e2}>^{C}$ is noticeably 
smaller in the case of 
$\nu_e \rightarrow \nu_s$ 
transitions of the solar neutrinos
than in the case of the 
$\nu_e \rightarrow \nu_{\mu(\tau)}$ transitions
considered above (for illustration see the 
corresponding plots for the 
Super-Kamiokande detector in \cite{SK98III}). 
For $\sin^22\theta_V = 0.01$, for instance,
the maximal values reached by 
$<P_{e2}>^{C}$ in the two cases are $\sim 0.1$ and $\sim 0.3$,
respectively. Similar result was shown \cite{SK98III} 
to be valid for $<P_{e2}>^{C}$ calculated
for the location of the Super-Kamiokande detector.
The explanation of this difference can be found in
\cite{ChPet992}: it is related to the 
fact that the total neutrino conversion
point at $\sin^22\theta_V \lesssim 0.25$,
which determines the enhancement under discussion,
takes place for the different $\hat{h}$
at $\sin^22\theta_V \cong (0.12 - 0.25)$,
which is relatively ``far'' in 
$\sin^22\theta_V$ from the region of interest
(see Table 4 in \cite{ChPet992}).
The primary source is, of course, the difference between
the neutrino effective potentials 
in matter in the two cases \cite{Lang87} 
(see also, e.g., \cite{SK98III}).

 For  $\sin^22\theta_V \lesssim (0.009 - 0.010)$,
the probability D-N asymmetries for the 
{\it Core} and {\it Night} event samples
have (as functions of $E/\Delta m^2$)
substantial negative parts  
in addition to the positive ones.
For most of the indicated values of 
$\sin^22\theta_V$ and the values of
$\Delta m^2$ 
from the MSW $\nu_e \rightarrow \nu_s$ solution 
region and of the neutrino energy of interest,
$E \geq 6.44$ MeV,
the energy-integrated 
D-N asymmetries $A^{C}$ and $A^{N}$
are determined essentially 
by the negative parts of 
$A^{C}$ and $A^{N}$ and therefore are
negative. 
The zero asymmetry regions take place 
for $A^{C}$ and $A^{N}$ roughly at
$\sin^22\theta_V \sim 0.009$ and 
$\sin^22\theta_V \sim 0.010$, respectively.
At small mixing angles,
$\sin^22\theta_V \lesssim 0.010$, the CC asymmetries are 
relatively small. 
% For $\sin^22\theta = 0.003;~0.005; 0.007;~0.010$, 
% for instance, $A_{P}^{C}$ and $A_{P}^{N}$ 
% have values in the intervals
% $((-2.3) - 0.8)\%;~((-3.0) - 2.4)\%;~((-3.0) - 6.0)\%;~
% ((-2.4) - 8.0)\%$ and
% $((-0.5) - 0.7)\%;~((-0.8) - 2.0)\%;~((-0.8) - 4.0)\%;~
% ((-0.7) - 6.0)\%$, respectively.
For $\sin^22\theta_V = (0.003 - 0.010)$, 
for instance, one finds 
$ - 3.0\% \lesssim A_{P}^{C} \lesssim 8.0\%$ 
and $ - 0.8\% \lesssim A_{P}^{N} \lesssim 6.0\%$. 

  The solar neutrino data strongly 
disfavors large mixing angle
MSW solution in the case of 
$\nu_e \rightarrow \nu_s$
transitions. Nevertheless, we would like 
to note that due to
the existence of points of absolute minima of 
the probability $P_{e2}$ for the solar 
neutrinos crossing the Earth core \cite{ChPet992},
$\mathrm{min}(P_{e2}) = 0$, one finds at 
$\sin^22\theta \sim (0.4 - 0.5)$ a large negative  
{\it Core} probability asymmetry,
$\mathrm{min}(A_{P}^{C}) \cong -70\%$, at 
$E/\Delta m^2 \sim (2 - 3)\times 10^{6}{\rm MeV/eV^2}$.
 
\vskip 0.3cm
\leftline{\bf 2.2.3 On the Latitude Dependence of the {\it Core} Asymmetry 
Enhancement}
\vskip 0.3cm

   The comparison of $<P_{e2}>^{C} = <P_{e2}>^{C}(\rho_r)$ 
plots at fixed
$\sin^22\theta_V \lesssim 0.01$ 
for the \SK\ detector
\cite{SK97I,SK98III}, where
\begin{equation}
\rho_r \equiv {{\Delta m^2 \cos2\theta_V}\over{\sqrt{2}G_{F}E}}m_{N},
\end{equation}
\noindent $m_{N}$ being the nucleon mass, 
with those for the SNO detector shows that the
main effect on $<P_{2e}>^{C}$
of a moderated increase 
of the SNO detector latitude  
is a shift of the position of the dominating
absolute maximum of $<P_{2e}>^{C}$ 
and not a reduction of this maximum. Thus, 
the magnitude of the {\it mantle-core enhancement} for
moderate increase of the detector latitude, and therefore the
sensitivity of the SNO detector to the
D-N effect, is primarily affected
by the reduction in the {\it Core} residence time 
and by changes in the
position of the $<P_{e2}>^{C}$ dominating maximum
along the $\rho_r$ axis with respect to the 
$\bar{P}_{\odot}(\nu_e \rightarrow \nu_e) =
1/2$ line. This observation
suggests that for SNO, ICARUS and  
other solar neutrino
detectors situated
at latitudes which are 
somewhat larger than
that of the Super-Kamiokande detector,
the sensitivity to the D-N effect may be
better than is usually expected.

  To understand this result 
it is helpful to consider 
the dependence of $P_{e2}$  
on $\rho_r$ and the Nadir angle 
$\hat{h}$. The behavior of
$<P_{e2}>^{C}$ follows from this dependence. 
At small $\sin^22\theta_V$
and for any fixed value of $\rho_r$ (or $E/\Delta m^2$), \PeTw\ is a
decreasing function of \hath\ for the Earth-core-crossing neutrinos
\footnote{This is related to the fact that 
with the increase of \hath\
the position of the 
total neutrino conversion 
point where $P_{e2} = 1$,
which determines the magnitude 
of the mantle-core enhancement
of $P_{e2}$ at $\sin^22\theta_V \lesssim 0.01$, 
shifts to larger 
values of $\sin^22\theta$ \cite{ChPet992}, e.g., from
$\sin^22\theta \cong 0.04$ at $\hat{h} = 13^{\circ}$ 
to $\sin^22\theta \cong 0.08$ at $\hat{h} = 30^{\circ}$
in the $\nu_e \rightarrow \nu_{\mu (\tau)}$
case. The indicated shift is caused, in particular,
by the decreasing of the neutrino path length
through the Earth core 
and of the average density of 
the core along the neutrino path
when $\hat{h}$ increases.}. 
Since the time averaged \hath\ for 
the core-crossing neutrinos
increases with the increasing of the detector 
latitude and given the fact that
SNO is at a higher latitude than the 
Super-Kamiokande detector, a smaller 
$<P_{e2}>^{C}$
probability is expected for SNO than for the 
Super-Kamiokande detector, which seems to be 
in contradiction with the suggestion made above.
A more detailed study of the $\rho_r$ and $\hat{h}$
dependence of $P_{e2}$ at fixed $\sin^22\theta_V \lesssim 0.01$
reveals, however, that in the region of the
absolute maximum of $P_{2e}$ (at a given $\sin^22\theta_V$)
this dependence can be described 
as a part of a {\it positive saddle}
with fastly decreasing ``shoulders'', but whose main axis
(i.e., the ``ridge'' or minimal gradient line) 
is curved towards smaller values of $\rho_r$
when $\hat{h}$ increases (Fig. 3). 
Moreover, along its main axis and 
on a relatively large part of it, the {\it saddle} is
nearly flat and therefore
$P_{2e}$ changes very little. 
The plots of $P_{2e}$ versus $\hat{h}$ 
at fixed $\rho_r$ are sections of
the {\it saddle} at high inclination with 
respect of the {\it saddle
axis}, so what they illustrate
is the fast decrease of the saddle shoulders.
The plots of the $P_{2e}$  dependence on 
$\rho_r$ at fixed but different $\hat{h}$
show the shift of the position of the 
maximum along the $\rho_r-$axis 
for changing $\hat{h}$, thus 
exhibiting the curvature 
of the saddle axis. 
The behavior of $P_{2e}$ discussed above
is illustrated in Fig. 3. 

   When translating these considerations into 
total event rate predictions, one obviously  
has to take into account 
the effects of the shape of the 
$^{8}B$ neutrino spectrum and
of the integration over the 
neutrino energy. For $\sin^22\theta_V$ 
in the range of $\sim (10^{-3} - 10^{-2})$, the D-N effect
is suppressed by the
simultaneous  presence of a negative and positive 
parts of the D-N probability asymmetry in 
the interval of values of $E/\Delta m^2$ 
of interest \cite{SK97I}.
As already discussed above, one of the 
important parameters 
which determine for given $\sin^22\theta_V$ 
the magnitude of the
D-N effect is the position of
the absolute maximum of $<P_{2e}>^{C}$ 
along the $E/\Delta m^2$ axis 
with respect to the 
zero asymmetry line $E/\Delta m^2 = (E/\Delta m^2)_{0}$
corresponding to $\Ps = 1/2$.
The shift towards smaller values of 
$\rho_r$ (or larger values of $E/\Delta m^2$) 
of the position of the dominating maximum 
of $<P_{2e}>^{C}$ as the
detector latitude increases 
makes the crossing of the 
$\Ps = 1/2$ line by the maximum occur at 
larger values of $\sin^22\theta_V$.
Thus, for a given $\sin^22\theta_V \lesssim 0.009$,
the region of negative values of 
$A_{P}^{C}$ in $E/\Delta m^2$ increases somewhat
with the increase of $\hat{h}$ at the expense of the  
region where $A_{P}^{C} > 0$. This results
in a certain decreasing of 
the CC total event rate {\it Core} 
D-N asymmetry $A_{D-N}^{C}$. 
Similar effect takes place for the {\it Night} 
asymmetry $A_{D-N}^{N}$.
The decreasing is 
roughly the same in magnitude as the one caused
by the presence of the NC contribution due to the
$\nu_{\mu(\tau)}$ in the Super-Kamiokande
solar neutrino signal in the case of the 
MSW $\nu_e \rightarrow \nu_{\mu(\tau)}$ transitions
of solar neutrinos. 
Actually, the ratio 
$A_{D-N}^{C}/A_{D-N}^{N}$ of the discussed 
asymmetries for SNO is typically 
bigger, as we shall see,
than that for the Super-Kamiokande detector.
This analysis shows also that 
in the case of the 
MSW SMA $\nu_e \rightarrow \nu_{\mu(\tau)}$ solution
the enhancement of the {\it Core}
D-N asymmetry in the CC event rates
for detectors located 
at different latitudes not exceeding,
say, the SNO latitude 
($\SNOLat = 46\deg \, 20'$\ North), 
is approximately
the same in magnitude 
and that the sensitivity to the
{\it Core} D-N effect, apart from the systematic
error, is limited essentially by 
the statistics of the corresponding data samples. 

\vskip 0.3cm
\leftline{\bf 3. Predictions for the Energy-Integrated 
D-N Asymmetries}
\vskip 0.3cm

  In this Section we will discuss the 
predictions we have obtained 
for the values of the various energy-integrated, 
one year averaged
D-N asymmetries for the SNO detector. 
The asymmetries are defined as in eq. (2) and
in \cite{SK97I,SK97II,SK98III}. We consider the 
following set of D-N asymmetries:
i) {\it Night}, {\it Core} and {\it Mantle} 
asymmetries in the CC
signal in SNO, 
$A_{D-N}^{N,C,M}(CC) = A_{D-N}^{N,C,M}(CC;~T_{e,th},Y_e^{c})$, 
in the case of the  
$\nu_e \rightarrow\nu_{\mu (\tau)}$ and
$\nu_e \rightarrow\nu_{s}$ transitions
of solar neutrinos,
ii) {\it Night}, {\it Core} and {\it Mantle} 
asymmetries in the ES
signal in SNO, 
$A_{D-N}^{N,C,M}(ES) = A_{D-N}^{N,C,M}(ES;~T_{e,th},Y_e^{c})$, 
in the case of the  
$\nu_e \rightarrow\nu_{\mu (\tau)}$ and
$\nu_e \rightarrow\nu_{s}$ transitions, and
iii) {\it Night}, {\it Core} and {\it Mantle}
asymmetries in the NC
signal in SNO, $A_{D-N}^{N,C,M}(NC) = A_{D-N}^{N,C,M}(NC;~Y_e^{c})$, 
in the case of the  
$\nu_e \rightarrow\nu_{s}$ transitions
\footnote{Note that with the definition 
of the D-N asymmetry
we use \cite{SK97I,SK97II,SK98III},
the one year average energy-integrated
D-N asymmetries we consider
are zero in the case of massless
standard model neutrinos.}.
The asymmetries $A_{D-N}^{N,C,M}(CC;~T_{e,th},Y_e^{c})$
and $A_{D-N}^{N,C,M}(ES;~T_{e,th},Y_e^{c})$
have been calculated for two values of
$T_{e,th}$: 5.0 MeV and 7.5 MeV. Results for
all  asymmetries for 
$Y_e(core) = 0.467~and~0.500$ \cite{SK97II} 
are given 
\footnote{Predictions for the {\it Night}
CC asymmetry $A_{D-N}^{N}(CC;~T_{e,th},Y_e^{c})$
for $T_{e,th}= 5$  and  $T_{e,th}= 5.0;~7.5~$ MeV
for the standard values of $Y_e(man)$ and 
$Y_e(core)$ were given respectively 
in \cite{LisiM97,BK97}
and \cite{BKS00}. Results for the {\it Night}
asymmetry in the NC data $A_{D-N}^{N}(NC)$
for the standard values of $Y_e(man)$ and 
$Y_e(core)$ were obtained also in
\cite{BKS00}.}.

   Our results are presented in Tables II - X 
and in Figs. 4a - 11b. The tables contain values 
of the three types of asymmetries
and of the ratio of the {\it Core} and {\it Night}
asymmetries, calculated for 36 pairs of values of
$\Delta m^2$ and $\sin^22\theta_V$ 
distributed evenly in the ``conservative'' 
MSW solution regions (older and most recent). 
Some of these pairs of values, 
as like those with 
$\sin^22\theta_V \gtrsim 0.03$ in the case of 
MSW $\nu_e \rightarrow \nu_s$ solution in 
Tables VI - X, have been
excluded by the recent Super-Kamiokande 
data, but we kept them for giving 
an idea about the magnitude of
the asymmetries corresponding to them. 
In Figs. 4a - 11b
contours of constant (D-N) asymmetries
in the $\Delta m^2 - \sin^22\theta_V$ plane are shown
(iso-(D-N) asymmetry contour plots) 
together with the ``conservative''
MSW solution regions from \cite{PK99}. 
We did not include 
in the present work the iso-(D-N) asymmetry
contour plots for the different (CC, ES, NC)
{\it Mantle} asymmetries because the latter
are relatively small and thus difficult to observe
\footnote{These plots can be found at 
the site: {\tt
http://www.pv.infn.it/$\sim$maris/nue/sno/index.html}.}.
Below we comment briefly our results for the
indicated D-N asymmetries.
 
\vskip 0.3cm
\leftline{\bf 3.1. Magnitude of the Asymmetries and their  
$Y_e(core)$ and $T_{e,th}$ Dependence}
\vskip 0.3cm
\leftline{\bf 3.1.1. $\nu_e \rightarrow \nu_{\mu(\tau)}$ Transitions}
\vskip 0.3cm
\leftline{\bf A. Asymmetries in the CC Event Sample}
\vskip 0.3cm
 
   The most remarkable feature of the D-N 
asymmetries under discussion in this case is the 
relatively large absolute values of the 
{\it Core} CC asymmetry
$A_{D-N}^{C}(CC)$ in most of the region
MSW SMA solution, as well as the strong 
{\it mantle-core enhancement} 
of $A_{D-N}^{C}(CC)$ 
% \cite{SPPLB43498,ChPet991,ChPet992}
with respect to the 
{\it Night} and {\it Mantle} asymmetries
in this region (Tables II and III).
Due to the indicated enhancement
we typically have 
$|A_{D-N}^{C}(CC)| > 1\%$ 
at $\sin^22\theta_V \cong (0.001 - 0.006)$
and $\Delta m^2 \sim (3.0 - 8.0)\times 10^{-6}~{\rm eV^2}$,
while the {\it Night} and {\it Mantle} asymmetries 
$|A_{D-N}^{N,M}(CC)| < 1\%$ and are hardly observable.
The asymmetry $A_{D-N}^{C}(CC)$ is negative and 
greater than 1\% in absolute value for 
$\sin^22\theta_V \cong (0.7 - 4.0)\times 10^{-3}$
and $\Delta m^2$ in the interval above,
reaching the minimal value of $\sim (-3.2)\%$
at $\sin^22\theta_V \sim 0.002$.
The ratio $|A_{D-N}^{C}(CC)/A_{D-N}^{N}(CC)|$ varies with
$\sin^22\theta_V$ and $\Delta m^2$ in the 
MSW SMA solution region
typically in the interval of $\sim (5 - 8)$.
For $\sin^22\theta_V \cong (0.008 - 0.010)$,
we have $A_{D-N}^{C}(CC) \cong (12.9 - 39.7)\%$, while
$A_{D-N}^{N}(CC) \cong (1.8 - 6.6)\%$, the ration 
between the two asymmetries varying in the range
of $\sim (4.3 - 7.9)$.
Actually, the CC {\it Night} asymmetry
$A_{D-N}^{N}(CC)$ is hardly observable
at $\sin^22\theta_V < 0.008$. The same conclusion 
is valid for the {\it Mantle} asymmetry 
$A_{D-N}^{M}(CC)$ at $\sin^22\theta_V < 0.010$.

    The {\it mantle-core enhancement} 
of the CC {\it Core} asymmetry 
% \cite{SPPLB43498,ChPet991,ChPet992} 
is rather modest in the region of the MSW LMA solution,
where we typically have 
$A_{D-N}^{N}(CC)/A_{D-N}^{N}(CC) \sim (1.0 - 1.3)$.
Nevertheless, both asymmetries 
$A_{D-N}^{C}(CC)$ and $A_{D-N}^{N}(CC)$
have values exceeding 1\% in most 
of the solution's region (Tables II and III).
More specifically, for 
$\Delta m^2 \lesssim 9.0\times 10^{-5}~{\rm eV^2}$,
we have $A_{D-N}^{N}(CC;~7.5~{\rm MeV}) \gtrsim 2.5\%$
and $A_{D-N}^{C}(CC;~5.0~{\rm MeV}) \gtrsim 5.0\%$.
These predictions suggest that the SNO experiment
will be able to probe most of the LMA solution region
by measuring the {\it Night} and/or {\it Core}
D-N asymmetries in the CC event sample.
The conclusion is robust against the change of 
$Y_e(core)$ in the interval 0.467 - 0.500.

   Changing $Y_e(core)$ from 0.467 to 0.500
affects little the {\it Night} CC asymmetry,
but can change noticeably, 
typically by $\sim (10 - 30)\%$,
the {\it Core} asymmetry in the SMA solution region
(see, e.g., cases 13 - 19 in Table II and 12-14 and 16 
in Table III).
For  $\sin^22\theta_V \sim (0.004 - 0.010)$
and $\Delta m^2 \sim (6.5 - 10.0)\times 10^{-6}~{\rm eV^2}$,
the change of $A_{D-N}^{C}(CC)$ 
can be quite dramatic - the asymmetry 
increases approximately by a factor of $\sim 1.5$.
Thus, for values of
$\sin^22\theta_V$ and $\Delta m^2$ 
from a specific subregion of the 
MSW SMA solution region the 
magnitude of the predicted  
{\it Core} asymmetry $A_{D-N}^{C}(CC)$ 
is very sensitive to the value of the 
electron fraction number $Y_e(core)$
and the average matter density 
of the Earth core. The former is determined
by the chemical composition of the core
\cite{YeEarth} (see also \cite{SK97II}).
If the MSW SMA solution turned out to be the
solution of the solar neutrino problem and
the parameters $\sin^22\theta$ and $\Delta m^2$ 
lied in the region specified above and are known
with a sufficiently high precision,
the measurement of the CC {\it Core} asymmetry
can give unique information, e.g., about the 
chemical composition of the Earth core.

  Increasing the value of the threshold kinetic energy
of the final state electron $T_{e,th}$
in the CC reaction, eq. (3), from 5.0 MeV to 7.5 MeV
can have a dramatic effect on
the {\it Core} asymmetry  
$A_{D-N}^{C}(CC)$ changing it by a factor of $\sim 2$
(compare cases 11, 18 and 20 in Tables II and III).
For the values of 
$\sin^22\theta_V$ and $\Delta m^2$ for which
the {\it Night} asymmetry $A_{D-N}^{N}(CC;~5~{\rm MeV}) > 1\%$,
the asymmetry for 
$T_{e,th} = 7.5$ MeV, $A_{D-N}^{N}(CC;~7.5~{\rm MeV})$,
can be both larger or smaller than 
$A_{D-N}^{N}(CC;~5~{\rm MeV})$ by at most $\sim 20\%$.
For given $\sin^22\theta_V$ and $\Delta m^2$ from the
MSW solution regions there should exist
optimal value of $T_{e,th} \gtrsim 5$~MeV for which
$|A_{D-N}^{C}(CC;~T_{e,th})|$ and/or
$|A_{D-N}^{N}(CC;~T_{e,th})|$ have maximal values.

   All the above properties of the
{\it Night} and {\it Core} CC 
asymmetries under discussion are exhibited clearly in the 
corresponding iso-(D-N) asymmetry contour plots
for $A_{D-N}^{N}(CC)$ and $A_{D-N}^{C}(CC)$,
shown in Figs. 4a-4b and 5a-5b, respectively.
% Figures 4a (4b) and 5a (5b) 
% correspond to $T_{e,th} = 5~(7.5)~{\rm MeV}$.
% In each of these figures results for 
% $Y_e(core) = 0.467;~0.500$ are presented.

\vskip 0.3cm
\leftline{\bf B. ES Event Sample Asymmetries}
\vskip 0.3cm

   The D-N asymmetries in the ES sample are typically smaller
by a factor of $\sim 2$ than the asymmetries in the CC sample
(compare Tables II, III with IV, V).
There are exceptions when  
$A_{D-N}^{C(N)}(ES)$ is of the 
order of or even larger than
$A_{D-N}^{C(N)}(CC)$ but not by factors 
which can compensate for the much lower
statistics expected in the ES sample of SNO.
The {\it mantle-core enhancement} of the {\it Core} asymmetry
at $\sin^22\theta \lesssim 0.01$
% \cite{SPPLB43498,ChPet991,ChPet992} 
is operative in this case as well:
one typically has 
$|A_{D-N}^{C}(ES)/A_{D-N}^{N}(ES)| \sim (5 - 8)$
(see Tables IV and V). 

     The above features of the 
ES sample asymmetries are exhibited in Figs. 6a - 6b and
7a - 7b, representing the iso-(D-N) asymmetry
contour plots for $A_{D-N}^{N}(ES)$
and $A_{D-N}^{C}(ES)$, respectively,
derived for $Y_e(core) = 0.467;~0.500$ and  
for $T_{e,th} = 5~{\rm MeV}~(a)$
and $T_{e,th} = 7.5~{\rm MeV}~(b)$.

\vskip 0.3cm
\leftline{\bf 3.1.2. $\nu_e \rightarrow \nu_{s}$ Transitions}
\vskip 0.3cm

  The predicted D-N asymmetries of the same type 
({\it Core}, {\it Night} or {\it Mantle}) 
in the CC, ES and NC
event samples do not 
differ much in magnitude 
in the case of the MSW $\nu_e \rightarrow \nu_{s}$
transition solution of the solar neutrino problem
(Tables VI - X).
Let us remind the reader that the observation of a
D-N asymmetry in the NC sample of events would
be a proof that solar neutrinos 
undergo transitions into sterile 
neutrinos \cite{KLPSter96}:
this asymmetry is zero in the case of the 
$\nu_e \rightarrow \nu_{\mu(\tau)}$ transitions.
The main contribution to the energy-integrated
D-N asymmetries in the three event samples 
comes, as in the case of the 
Super-Kamiokande detector \cite{SK98III},
from the ``high'' energy tail
of the $^{8}$B neutrino spectrum.
This is related to the fact the relevant 
neutrino effective potential difference
in the Earth matter in the case of 
the $\nu_e \rightarrow \nu_{s}$ transitions
is approximately by a factor of 2 smaller 
than in the case of 
$\nu_e \rightarrow \nu_{\mu(\tau)}$
transitions \cite{Lang87,SK98III}.
Only the {\it Core} asymmetries 
can exceed $1\%$ in absolute value
at $\sin^22\theta_V \lesssim 0.01$
in specific limited
subregions of the ``conservative'' 
MSW solution region. 
The {\it Night} asymmetries in the 
``conservative'' solution region are
for all samples smaller than 1\% for
$\sin^22\theta_V \lesssim 0.01$.
In the subregion where 
$\sin^22\theta_V > 0.01$
both the {\it Core} and {\it Night}
asymmetries for the three event samples,
$A_{D-N}^{C,N}(CC)$,
$A_{D-N}^{C,N}(ES)$ and 
$A_{D-N}^{C,N}(NC)$,
are greater than 1\%,
the {\it Core} asymmetries being larger than
the corresponding {\it Night}
asymmetries by a factor of
$\sim (2 - 5)$.

   The above features can be seen in
the corresponding iso-(D-N) asymmetry contour plots
shown in Figs. 8a - 8b (CC {\it Night} asymmetry),
9a - 9c (CC {\it Core} asymmetry),
10a - 10b (ES {\it Night} and {\it Core} asymmetries)
and 11a - 11b (NC {\it Night} and {\it Core} asymmetries).    

   Let us note finally that in the subregion
$\Delta m^2 \cong (4.0 - 7.0)\times 10^{-6}~{\rm eV^2}$,
$\sin^22\theta \cong (0.009 - 0.020)$, of the 
``conservative''solution region,
the CC {\it Core} asymmetry
$A_{D-N}^{C}(CC)$ exhibits a very
strong dependence on
the value of $Y_e(core)$: the change of 
$Y_e(core)$ from the value of 0.467 to
0.500 leads to an increase of 
$A_{D-N}^{C}(CC)$ by a factor of $\sim (2 - 4)$
(see cases 26 - 33 in Tables VI and VII). This becomes evident
also by comparing Figs. 9a and 9b.

\vskip 0.3cm
\leftline{\bf 4. Conclusions}
\vskip 0.3cm
 In the present article we have derived 
detailed predictions for the day-night (D-N) asymmetry 
in the energy-integrated one year signals in the SNO detector
in the case of the MSW
$\nu_e \rightarrow \nu_{\mu(\tau)}$
and/or $\nu_e \rightarrow \nu_{s}$ transition solutions
of the solar neutrino problem. 
The asymmetries in the
charged current (CC), eq. (3), and 
$\nu - e^{-}$ elastic scattering (ES),
eq. (4), event rates were calculated 
for both MSW solutions;
in the case of the $\nu_e \rightarrow \nu_{s}$ 
transition solution
the D-N asymmetry in the neutral current 
(NC) event rate, eq. (5),
was derived as well.
The asymmetries were calculated for 
the three night samples of events which are produced by
the solar neutrinos crossing i) the Earth
mantle only ({\it Mantle}), ii) the Earth core ({\it Core}) 
and iii) the Earth core and/or the mantle ({\it Night})
on the way to the SNO dtector: $A_{D-N}^{C,N,M}(CC)$,
$A_{D-N}^{C,N,M}(ES)$ and 
$A_{D-N}^{C,N,M}(NC)$.

   We have studied the effects of the uncertainties
in the values of the cross-sections of the CC and NC
neutrino-induced reactions on deuterium
by comparing the predictions for the asymmetries
$A_{D-N}^{C,N,M}(CC)$ and $A_{D-N}^{C,N,M}(NC)$,
obtained using the CC cross-sections from refs.
\cite{XS:IAUTH0,XS:IAUTH1,XS:IAUTH2},
and the NC cross-sections from refs.
\cite{XS:IAUTH1,XS:IAUTH2}. 
Our analysis showed that 
even neglecting
the electron energy resolution function and the
expected errors in the measurements, 
the calculation of the D-N asymmetries in the 
CC and NC event samples in the SNO detector 
utilizing the different 
cross-sections from
\cite{XS:IAUTH0,XS:IAUTH1,XS:IAUTH2}
and \cite{XS:IAUTH1,XS:IAUTH2}, respectively,
should give the same results within 1\% 
in the MSW ``conservative'' solution regions.

  As in the case of the same type of
D-N asymmetries for the
Super-Kamiokande detector \cite{SK97I,SK97II,SK98III},
at small mixing angles the 
{\it Core}  asymmetries were shown to be 
considerably larger than the corresponding 
{\it Night} and {\it Mantle} asymmetries  
due to the {\it mantle-core enhancement} effect
\cite{SPPLB43498,ChPet991,ChPet992}.
Due to this enhancement, e.g.,
the one year average {\it Core} CC asymmetry
in the case of the SMA MSW 
$\nu_e \rightarrow \nu_{\mu(\tau)}$ solution
is bigger than 1\% in absolute value, 
$|A_{D-N}^{C}(CC)| > 1\%$, 
% at $\sin^22\theta_V \cong(0.7 - 6.0)\times 10^{-3}$
% in a  rather large $\Delta m^2$ subregion,
% $\Delta m^2 \sim (3.0 - 8.0)\times 10^{-6}~{\rm eV^2}$.
in large subregions of the ``conservative'' solution region
which extends in $\sin^22\theta_V$ from  
$\sim 0.7\times 10^{-3}$ to $\sim 10^{-2}$
(Figs. 5a - 5b).
In contrast, in the 
% ``conservative'' SMA 
same solution region
we have $|A_{D-N}^{N(M)}(CC)| < 1\%$
for the {\it Night} ({\it Mantle}) asymmetry 
at $\sin^22\theta_V < 0.008~(0.010)$ (Figs. 4a - 4b).
The asymmetry $A_{D-N}^{C}(CC)$ is negative and 
greater than 1\% in absolute value for 
$\sin^22\theta_V \cong (0.7 - 4.0)\times 10^{-3}$
and $\Delta m^2\sim (3.0 - 8.0)\times 10^{-6}~{\rm eV^2}$, 
% indicated above,
reaching the minimal value of $\sim (-3.2)\%$.
% at $\sin^22\theta_V \sim 0.002$.
The ratio $|A_{D-N}^{C}(CC)/A_{D-N}^{N}(CC)|$ varies with
$\sin^22\theta_V$ and $\Delta m^2$ in the 
MSW SMA solution region
typically in the interval of $\sim (5 - 8)$
(Tables II and III).
For $\sin^22\theta_V \cong (0.008 - 0.010)$,
we have in this region
$A_{D-N}^{C}(CC) \cong (12.9 - 39.7)\%$, while
$A_{D-N}^{N}(CC) \cong (1.8 - 6.6)\%$.

   In the region of the MSW LMA
$\nu_e \rightarrow \nu_{\mu(\tau)}$ solution
we typically have 
$A_{D-N}^{N}(CC)/A_{D-N}^{N}(CC) \sim (1.0 - 1.3)$.
Both asymmetries 
$A_{D-N}^{C}(CC)$ and $A_{D-N}^{N}(CC)$
have values exceeding 1\% in most 
of the solution region (Tables II, III and Figs. 4a - 5b).

    The D-N asymmetries in the ES sample 
in the case of the MSW 
$\nu_e \rightarrow \nu_{\mu(\tau)}$ solution
are in most of the ``conservative'' solution regions smaller
by a factor of $\sim 2$ than the asymmetries in the CC sample
(Tables II, III and IV, V, Figs. 6a - 7b).
There are exceptions when  
$A_{D-N}^{C(N)}(ES)/A_{D-N}^{C(N)}(CC)\gtrsim 1$,
but the ratio factors involved 
cannot compensate for the much lower
statistics expected in the ES sample of SNO.

  The predicted D-N asymmetries of the same type 
({\it Core}, {\it Night} or {\it Mantle}) 
in the CC, ES and NC
event samples do not 
differ much in magnitude 
in the case of the MSW $\nu_e \rightarrow \nu_{s}$
transition solution of the solar neutrino problem
(Tables VI - X, Figs. 8a - 11b).
Only the {\it Core} asymmetries 
can exceed $1\%$ in absolute value
at $\sin^22\theta_V \lesssim 0.01$
in specific limited
subregions of the ``conservative''
solution region. 
In the subregion where 
$\sin^22\theta_V > 0.01$
both the {\it Core} and {\it Night}
asymmetries for the three event samples,
are greater than 1\%,
the {\it Core} asymmetries exceeding
the corresponding {\it Night}
asymmetries by a factor of
$\sim (2 - 5)$. 

     We have found also that 
in certain subregions of the MSW
solution regions at small $\sin^22\theta_V$,
the predicted magnitude
of the {\it Core} D-N asymmetry
in the CC and ES samples is very sensitive
to the value of the electron fraction number
$Y_e^{c}$  (i.e., to the chemical composition)
and to the average matter density 
of the Earth core:
the asymmetry $A_{D-N}^{C}(CC)$, for instance,
can increase in the case of the 
$\nu_e \rightarrow \nu_{\mu (\tau)}$
($\nu_e \rightarrow \nu_{s}$) solution
by a factor of $\sim 1.5~(2-4)$
when $Y_e^{c}$ is changed from 
the standard value of 0.467 \cite{Stacey77,PREM81} 
to its possibly conservative upper limit
of 0.500 \cite{YeEarth,SK97II} (Tables II, III, VI, VII,
Figs. 5a, 5b, 7a, 7b, 9a-9b).  

 We have studied the dependence of the D-N asymmetries
considered on the final state e$^{-}$ threshold energy
$T_{e,th}$ in the
CC and ES reactions. Increasing the value of $T_{e,th}$
in the CC and ES reactions from 5.0 MeV to 7.5 MeV
can have a dramatic effect on
the {\it Core} asymmetries
$A_{D-N}^{C}(CC)$ and $A_{D-N}^{C}(ES)$,  
changing, e.g., $A_{D-N}^{C}(CC)$ by a factor of $\sim 2$
in the case of the $\nu_e \rightarrow \nu_{\mu (\tau)}$
solution (Tables II, III, VI, VII, Figs. 5a, 5b, 9a - 9c).

 Our analysis showed also that 
for the MSW SMA 
$\nu_e \rightarrow \nu_{\mu(\tau)}$ 
and $\nu_e \rightarrow \nu_{s}$ 
solutions, the enhancement of the {\it Core}
D-N asymmetry in the CC event rates
for detectors located 
at different latitudes not exceeding,
say, the SNO latitude 
($\SNOLat = 46\deg \, 20'$\ North), 
is approximately
the same in magnitude.
This result
suggests that for SNO, ICARUS and  
other solar neutrino
detectors operating
at latitudes which are 
somewhat larger than
that of the Super-Kamiokande detector,
the sensitivity to the {\it Core} 
D-N effect, apart from the systematic
error, is limited essentially by 
the statistics of the 
corresponding data samples.

 The results of the present study allow us to 
conclude, in particular,
that the SNO experiment has the potential of
probing substantial parts of 
the SMA and LMA MSW 
$\nu_e \rightarrow \nu_{\mu(\tau)}$
solution regions by performing
{\it Night} and {\it Core}
D-N asymmetry measurements in the
CC event sample. The
measurements of the D-N effect related observables 
with the SNO detector appears unlikely 
to provide by itself a critical test of 
the MSW $\nu_e \rightarrow \nu_s$ (SMA)
solution of the solar neutrino problem.
It is expected that such a test 
will be provided by the
comparison of the CC and NC SNO data.
Nevertheless, certain subregions 
of the ``conservative'' 
MSW $\nu_e \rightarrow \nu_{s}$ solution region 
can also be probed through the measurements of the 
D-N effect, e.g., in the
CC and NC event samples.    

\vskip 0.3cm
\leftline{\bf Acknowledgements.}
\vskip 0.3cm 
We would like to thank H. Robertson for 
clarifying discussions concerning the SNO experiment. 
The work of (S.T.P.) was supported in part by 
the EC grant ERBFMRX CT96 0090.

\newpage

% \section*{Tables}

\begin{table}[ht]

\begin{center}

\caption{The predicted position of the maximum,
value at the maximum and F.W.H.M.
of the function $\calS(E) \sigma(E)~$ 
for the different reactions  used 
for the detection of the solar neutrinos at SNO. 
The maximal value for the CC reaction is
normalized to the predicted one 
in ref. [27] for
$T_{e,th} = 0~$ MeV, while the maximal value 
for ES reaction is normalized to $T_{e,th} = 0~$ MeV.}

\begin{tabular}{cccccc}
  &  &   &   &   Normalized & \\
 Reaction & Ref. & \TeTh & $E^{\mathrm{max}}$ & Maximum &
F.W.H.M. \\
    &   & (MeV) & (MeV) & Value & (MeV) \\
 \hline \hline
 CC & \cite{XS:IAUTH0} & 0.0 & 9.58 & 1.00 & 5.770 \\
 CC & \cite{XS:IAUTH0} & 5.0 & 9.84 & 0.94 & 4.744 \\
 CC & \cite{XS:IAUTH0} & 7.5 & 10.86 & 0.68 & 3.049 \\
 \hline
 CC & \cite{XS:IAUTH1} & 0.0 & 9.60 & 1.10 & 5.757 \\
 CC & \cite{XS:IAUTH1} & 5.0 & 9.86 & 1.04 & 4.750 \\
 CC & \cite{XS:IAUTH1} & 7.5 & 10.88 & 0.75 & 3.076 \\
 \hline
 CC & \cite{XS:IAUTH2} & 0.0 & 9.58 & 1.04 & 5.767 \\
 CC & \cite{XS:IAUTH2} & 5.0 & 9.86 & 0.98 & 4.758 \\
 CC & \cite{XS:IAUTH2} & 7.5 & 10.86 & 0.71 & 3.055 \\
 \hline \hline
 ES & \cite{ESCross} & 0.0 & 8.04 & 1.00 & 6.948 \\
 ES & \cite{ESCross} & 5.0 & 9.34 & 0.40  & 5.285 \\
 ES & \cite{ESCross} & 7.5 & 10.40 & 0.18 & 3.932 \\
 \hline

\end{tabular}

\end{center}
% \label{tab:CSTAB}
\end{table}

\newpage

\vspace{0.5cm}

 \begin{table}[ht]                                                                                                                      
 \begin{center}                                                                                                                         
 \caption{D - N Asymmetries for the SNO Detector 
for  $\TeTh = 5$\ MeV, CC, 
$\nue\rightarrow\numt$\ transition}\label{Tab:Asym:act:CC:5}
\begin{tabular}{|rrr||rrrr||rrrr|}                                                                                                     
 \hline                                                                                                                                 
  &                                                                                                                                     
 &                                                                                                                                      
 &                                                                                                                                      
 \multicolumn{4}{c||}{$\YeCore = 0.467$} &                                                                                              
 \multicolumn{4}{c|}{$\YeCore = 0.5$} \\                                                                                                
  &                                                                                                                                     
 &                                                                                                                                      
 &                                                                                                                                      
 \multicolumn{3}{c}{$\AsymRs \times 100$} &                                                                                             
 \multicolumn{1}{c||}{$\left|\frac{\AsymRC}{\AsymRN}\right|$} &                                                                         
 \multicolumn{3}{c}{$\AsymRs \times 100$} &                                                                                             
 \multicolumn{1}{c|}{$\left|\frac{\AsymRC}{\AsymRN}\right|$} \\
\hline                                                                                                                                 
 \end{tabular}                                                                                                                          
 \end{center}                                                                                                                           
 \end{table}

%%%%%%%%%%%%%%%

\newpage

%%%%%%%%%%%%%%%

 \begin{table}[ht]                                                                                                                          
 \begin{center}                                                                                                                             
 \caption{D - N Asymmetries for the SNO Detector for  $\TeTh = 7.5$\ MeV, CC, $\nue\rightarrow\numt$\ transition}\label{Tab:Asym:act:CC:7.5}
 \begin{tabular}{|rrr||rrrr||rrrr|}                                                                                                         
 \hline                                                                                                                                     
  &                                                                                                                                         
 &                                                                                                                                          
 &                                                                                                                                          
 \multicolumn{4}{c||}{$\YeCore = 0.467$} &                                                                                                  
 \multicolumn{4}{c|}{$\YeCore = 0.5$} \\                                                                                                    
  &                                                                                                                                         
 &                                                                                                                                          
 &                                                                                                                                          
 \multicolumn{3}{c}{$\AsymRs \times 100$} &                                                                                                 
 \multicolumn{1}{c||}{$\left|\frac{\AsymRC}{\AsymRN}\right|$} &                                                                             
 \multicolumn{3}{c}{$\AsymRs \times 100$} &                                                                                                 
 \multicolumn{1}{c|}{$\left|\frac{\AsymRC}{\AsymRN}\right|$} \\
\hline                                                                                                                                     
 \end{tabular}                                                                                                                              
 \end{center}                                                                                                                               
 \end{table}

%%%%%%%%%%%%%%%

\newpage

%%%%%%%%%%%%%%%

 \begin{table}[ht]                                                                                                                      
 \begin{center}                                                                                                                         
 \caption{D - N Asymmetries for the SNO Detector for  $\TeTh = 5$\ MeV, ES, $\nue\rightarrow\numt$\ transition}\label{Tab:Asym:act:ES:5}
 \begin{tabular}{|rrr||rrrr||rrrr|}                                                                                                     
 \hline                                                                                                                                 
  &                                                                                                                                     
 &                                                                                                                                      
 &                                                                                                                                      
 \multicolumn{4}{c||}{$\YeCore = 0.467$} &                                                                                              
 \multicolumn{4}{c|}{$\YeCore = 0.5$} \\                                                                                                
  &                                                                                                                                     
 &                                                                                                                                      
 &                                                                                                                                      
 \multicolumn{3}{c}{$\AsymRs \times 100$} &                                                                                             
 \multicolumn{1}{c||}{$\left|\frac{\AsymRC}{\AsymRN}\right|$} &                                                                         
 \multicolumn{3}{c}{$\AsymRs \times 100$} &                                                                                             
 \multicolumn{1}{c|}{$\left|\frac{\AsymRC}{\AsymRN}\right|$} \\
\hline                                                                                                                                 
 \end{tabular}                                                                                                                          
 \end{center}                                                                                                                           
 \end{table}

%%%%%%%%%%%%%%%

\newpage

%%%%%%%%%%%%%%%

 \begin{table}[ht]                                                                                                                          
 \begin{center}                                                                                                                             
 \caption{D - N Asymmetries for the SNO Detector for  $\TeTh = 7.5$\ MeV, ES, $\nue\rightarrow\numt$\ transition}\label{Tab:Asym:act:ES:7.5}
 \begin{tabular}{|rrr||rrrr||rrrr|}                                                                                                         
 \hline                                                                                                                                     
  &                                                                                                                                         
 &                                                                                                                                          
 &                                                                                                                                          
 \multicolumn{4}{c||}{$\YeCore = 0.467$} &                                                                                                  
 \multicolumn{4}{c|}{$\YeCore = 0.5$} \\                                                                                                    
  &                                                                                                                                         
 &                                                                                                                                          
 &                                                                                                                                          
 \multicolumn{3}{c}{$\AsymRs \times 100$} &                                                                                                 
 \multicolumn{1}{c||}{$\left|\frac{\AsymRC}{\AsymRN}\right|$} &                                                                             
 \multicolumn{3}{c}{$\AsymRs \times 100$} &                                                                                                 
 \multicolumn{1}{c|}{$\left|\frac{\AsymRC}{\AsymRN}\right|$} \\
\hline                                                                                                                                     
 \end{tabular}                                                                                                                              
 \end{center}                                                                                                                               
 \end{table}

%%%%%%%%%%%%%%%

\newpage

%%%%%%%%%%%%%%%

 \begin{table}[ht]                                                                                                                     
 \begin{center}                                                                                                                        
 \caption{D - N Asymmetries for the SNO Detector for  $\TeTh = 5$\ MeV, CC, $\nue\rightarrow\nus$\ transition}\label{Tab:Asym:ste:CC:5}
 \begin{tabular}{|rrr||rrrr||rrrr|}                                                                                                    
 \hline                                                                                                                                
  &                                                                                                                                    
 &                                                                                                                                     
 &                                                                                                                                     
 \multicolumn{4}{c||}{$\YeCore = 0.467$} &                                                                                             
 \multicolumn{4}{c|}{$\YeCore = 0.5$} \\                                                                                               
  &                                                                                                                                    
 &                                                                                                                                     
 &                                                                                                                                     
 \multicolumn{3}{c}{$\AsymRs \times 100$} &                                                                                            
 \multicolumn{1}{c||}{$\left|\frac{\AsymRC}{\AsymRN}\right|$} &                                                                        
 \multicolumn{3}{c}{$\AsymRs \times 100$} &                                                                                            
 \multicolumn{1}{c|}{$\left|\frac{\AsymRC}{\AsymRN}\right|$} \\
\hline                                                                                                                                
 \end{tabular}                                                                                                                         
 \end{center}                                                                                                                          
 \end{table}

%%%%%%%%%%%%%%%

\newpage

%%%%%%%%%%%%%%%

 \begin{table}[ht]                                                                                                                         
 \begin{center}                                                                                                                            
 \caption{D - N Asymmetries for the SNO Detector for  $\TeTh = 7.5$\ MeV, CC, $\nue\rightarrow\nus$\ transition}\label{Tab:Asym:ste:CC:7.5}
 \begin{tabular}{|rrr||rrrr||rrrr|}                                                                                                        
 \hline                                                                                                                                    
  &                                                                                                                                        
 &                                                                                                                                         
 &                                                                                                                                         
 \multicolumn{4}{c||}{$\YeCore = 0.467$} &                                                                                                 
 \multicolumn{4}{c|}{$\YeCore = 0.5$} \\                                                                                                   
  &                                                                                                                                        
 &                                                                                                                                         
 &                                                                                                                                         
 \multicolumn{3}{c}{$\AsymRs \times 100$} &                                                                                                
 \multicolumn{1}{c||}{$\left|\frac{\AsymRC}{\AsymRN}\right|$} &                                                                            
 \multicolumn{3}{c}{$\AsymRs \times 100$} &                                                                                                
 \multicolumn{1}{c|}{$\left|\frac{\AsymRC}{\AsymRN}\right|$} \\
\hline                                                                                                                                    
 \end{tabular}                                                                                                                             
 \end{center}                                                                                                                              
 \end{table}

%%%%%%%%%%%%%%%

\newpage

%%%%%%%%%%%%%%%

 \begin{table}[ht]                                                                                                                     
 \begin{center}                                                                                                                        
 \caption{D - N Asymmetries for the SNO Detector for  $\TeTh = 5$\ MeV, ES, $\nue\rightarrow\nus$\ transition}\label{Tab:Asym:ste:ES:5}
 \begin{tabular}{|rrr||rrrr||rrrr|}                                                                                                    
 \hline                                                                                                                                
  &                                                                                                                                    
 &                                                                                                                                     
 &                                                                                                                                     
 \multicolumn{4}{c||}{$\YeCore = 0.467$} &                                                                                             
 \multicolumn{4}{c|}{$\YeCore = 0.5$} \\                                                                                               
  &                                                                                                                                    
 &                                                                                                                                     
 &                                                                                                                                     
 \multicolumn{3}{c}{$\AsymRs \times 100$} &                                                                                            
 \multicolumn{1}{c||}{$\left|\frac{\AsymRC}{\AsymRN}\right|$} &                                                                        
 \multicolumn{3}{c}{$\AsymRs \times 100$} &                                                                                            
 \multicolumn{1}{c|}{$\left|\frac{\AsymRC}{\AsymRN}\right|$} \\
\hline                                                                                                                                
 \end{tabular}                                                                                                                         
 \end{center}                                                                                                                          
 \end{table}

%%%%%%%%%%%%%%%

\newpage

%%%%%%%%%%%%%%%

 \begin{table}[ht]                                                                                                                         
 \begin{center}                                                                                                                            
 \caption{D - N Asymmetries for the SNO Detector for  $\TeTh = 7.5$\ MeV, ES, $\nue\rightarrow\nus$\ transition}\label{Tab:Asym:ste:ES:7.5}
 \begin{tabular}{|rrr||rrrr||rrrr|}                                                                                                        
 \hline                                                                                                                                    
  &                                                                                                                                        
 &                                                                                                                                         
 &                                                                                                                                         
 \multicolumn{4}{c||}{$\YeCore = 0.467$} &                                                                                                 
 \multicolumn{4}{c|}{$\YeCore = 0.5$} \\                                                                                                   
  &                                                                                                                                        
 &                                                                                                                                         
 &                                                                                                                                         
 \multicolumn{3}{c}{$\AsymRs \times 100$} &                                                                                                
 \multicolumn{1}{c||}{$\left|\frac{\AsymRC}{\AsymRN}\right|$} &                                                                            
 \multicolumn{3}{c}{$\AsymRs \times 100$} &                                                                                                
 \multicolumn{1}{c|}{$\left|\frac{\AsymRC}{\AsymRN}\right|$} \\
\hline                                                                                                                                    
 \end{tabular}                                                                                                                             
 \end{center}                                                                                                                              
 \end{table}

%%%%%%%%%%%%%%%

\newpage

%%%%%%%%%%%%%%%

 \begin{table}[ht]                                                                                                
 \begin{center}                                                                                                   
 \caption{D - N Asymmetries for the SNO Detector for NC, $\nue\rightarrow\nus$\ transition}\label{Tab:Asym:ste:NC}
 \begin{tabular}{|rrr||rrrr||rrrr|}                                                                               
 \hline                                                                                                           
  &                                                                                                               
 &                                                                                                                
 &                                                                                                                
 \multicolumn{4}{c||}{$\YeCore = 0.467$} &                                                                        
 \multicolumn{4}{c|}{$\YeCore = 0.5$} \\                                                                          
  &                                                                                                               
 &                                                                                                                
 &                                                                                                                
 \multicolumn{3}{c}{$\AsymRs \times 100$} &                                                                       
 \multicolumn{1}{c||}{$\left|\frac{\AsymRC}{\AsymRN}\right|$} &                                                   
 \multicolumn{3}{c}{$\AsymRs \times 100$} &                                                                       
 \multicolumn{1}{c|}{$\left|\frac{\AsymRC}{\AsymRN}\right|$} \\
\hline                                                                                                           
 \end{tabular}                                                                                                    
 \end{center}                                                                                                     
 \end{table}                                                                                                      
 
\newpage 
%%%%%%%%%%%%%%%%%%%%%%%%%%%%%%%%%%%%%%%%%
% FIGURES CAPTIONS                      %
%%%%%%%%%%%%%%%%%%%%%%%%%%%%%%%%%%%%%%%%%

\begin{center} {\bf FIGURE CAPTIONS}\end{center}

%%%%%%%%%%%%%%%%%%%%%%%%%%%%%%%%%%%%%
% CAPTION OF FIGURES : a - adccc21a.ps   %
%  b - sdccc21a.ps; c - sdcnc21a.ps; 
% d -adfcc21a.ps; ; e - sdfcc21a.ps %
% f - sdfnc21a.ps  %
%%%%%%%%%%%%%%%%%%%%%%%%%%%%%%%%%%%%%
\noindent 
{\bf Figures 1a - 1f.}
The variation with $\Delta m^2$ and $\sin^22\theta_V$ 
of the difference between the values of the 
{\it Core} ({\it Night}) D-N asymmetry
(a) - (c) ((d) - (f)),
calculated utilizing the predictions for the 
relevant CC and NC reaction cross-sections
derived in ref. \cite{XS:IAUTH1} and in 
ref. \cite{XS:IAUTH2}:
figures (a), (b), (e)
((c), (f))  
show the asymmetry difference
in the one year averaged CC (NC) event
rate; figures (a), (d) ((b), (c), (e) and (f))
correspond to 
$\nu_e \rightarrow \nu_{\mu(\tau)}$ 
($\nu_e \rightarrow \nu_{s}$)
transitions. The light-grey spot-like regions  
in the upper left-hand panels 
correspond to asymmetry differences 
exceeding 1\%, while the dark-grey areas
are the regions of the MSW solutions. 

\vspace{0.3cm}
%%%%%%%%%%%%%%%%%%%%%%%%%%%%%%%%%%%%%
% CAPTION OF FIGURE : fig:pe2%
%%%%%%%%%%%%%%%%%%%%%%%%%%%%%%%%%%%%%

\noindent {\bf Figures 2a - 2b.}
The probabilities $<P_{e2}>^{C}$ (solid line), $<P_{e2}>^{N}$
(dashed line) and $<P_{e2}>^{M}$ (dash-dotted line)
as functions of  $E/\Delta m^2$ for 
$\sin^22\theta_V = 0.005~(a);~0.010~(b)$
in the case of 
$\nu_e\rightarrow \nu_{\mu(\tau)}$
% and $\nu_e \rightarrow \nu_s$
transitions of solar neutrinos.

\vspace{0.3cm}
%%%%%%%%%%%%%%%%%%%%%%%%%%%%%%%%%%%%%
% CAPTION OF FIGURE : fig:pe2:curve %
%%%%%%%%%%%%%%%%%%%%%%%%%%%%%%%%%%%%%

\noindent {\bf Figure 3.} 
%\ref{fig:pe2:curve}:
The dependence of $P_{e2}$  
on $\rho_r$ and the Nadir angle
$\hat{h}$ for $\sin^22\theta_V = 0.01$
in the case of MSW $\nu_e\rightarrow \nu_{\mu(\tau)}$
(upper left cannel) and
$\nu_e \rightarrow \nu_s$ (upper right cannel)
transitions of solar neutrinos. The 
grey-scales correspond to
different values of
$P_{e2}$, as is indicated 
in the two vertical columns between the two
upper panels. The thin solid lines in both 
upper panels represent contours
of constant $P_{e2}$ values:
$P_{e2} =0.01$, $0.025$, $0.05$, $0.1$,
$0.15$, $0.2$, $0.25$, $0.30$, $0.40$.
The dotted and the thick black
lines are minimum gradient axes,
connecting the points of local maxima of 
$P_{e2}$ in the variable $\rho_r$ at 
fixed $\hat{h}$ (``ridges''), 
while the dash-dotted lines  
are minimum gradient axes 
connecting points of local
minima of $P_{e2}$ (``valleys''). 
The solid lines in the lower panels 
show $P_{e2}$ as a function of $\hat{h}$,
computed along the ``ridge'' 
leading to the absolute maximum of $P_{e2}$ 
at $\hat{h} = 0^{\circ}$, 
while the dashed lines 
show the dependence of $P_{e2}$ 
on $\hat{h}$ along the
$\rho_r =~const.$ line
(dashed lines in the upper panels) 
starting from the point
of the absolute maximum at 
$\hat{h} = 0^{\circ}$. 

\vspace{0.3cm}
%%%%%%%%%%%%%%%%%%%%%%%%%%%%%%%%
% CAPTION OF FIGURE : ac0f50ab,
% ac0f75ab, ac0c50ab, ac0c75ab
%%%%%%%%%%%%%%%%%%%%%%%%%%%%%%%%

\noindent {\bf Figures 4a - 4b, 5a - 5b.}
Iso - (D-N) asymmetry contour plots 
for the one year average CC {\it Night} (4a,4b),
and {\it Core} (5a,5b) asymmetries
for the SNO detector in the case of the 
$\nu_e \rightarrow\nu_{\mu (\tau)}$ transitions 
and $\TeTh = 5.0~(a);~7.5~(b)$ MeV.
The solid (dashed) lines correspond to $\YeCore = 0.467~ 
(0.500)$. The MSW SMA and LMA 
``conservative'' solution regions from ref. \cite{PK99} 
are also shown.

\vspace{0.3cm}
%%%%%%%%%%%%%%%%%%%%%%%%%%%%%%%%
% CAPTION OF FIGURE : ae_f50ab,
% ae_f75ab, ae_c50ab, ae_c75ab
%%%%%%%%%%%%%%%%%%%%%%%%%%%%%%%%

\noindent {\bf Figures 6a - 6b, 7a - 7b.}
% Figure \ref{ac0f50ab}: 
Iso - (D-N) asymmetry contour plots 
for the one year average ES {\it Night} (6a,6b),
and {\it Core} (7a,7b) asymmetries 
for the SNO detector in the case of the 
$\nu_e \rightarrow\nu_{\mu (\tau)}$ transitions 
and $\TeTh = 5.0~(a);~7.5~(b)$ MeV.
The solid (dashed) lines correspond to $\YeCore = 0.467~
(0.500)$. The MSW SMA and LMA 
``conservative'' solution regions \cite{PK99} 
are also shown.

\vspace{0.3cm}
%%%%%%%%%%%%%%%%%%%%%%%%%%%%%%%%
% CAPTION OF FIGURE : sc0f50ab,
% sc0f75ab, sc0c50a;b, sc0c75a;b
%%%%%%%%%%%%%%%%%%%%%%%%%%%%%%%%

\noindent {\bf Figures 8a - 8b, 9a - 9c.}
% Figure \ref{ac0f50ab}: 
The same as in figures 4a - 4b and 5a - 5b
in the case of the 
$\nu_e \rightarrow\nu_{s}$ transitions
of solar neutrinos.
The CC {\it Core} asymmetry in figures
9a - 9c correspond to
$\TeTh = 5.0~(a,b);~7.5~(c)$
and $\YeCore = 0.467~(a,c);~0.500~(b)$. 
The MSW $\nu_e \rightarrow\nu_{s}$ transition  
solution regions from ref. \cite{PK99} 
are also shown (the ``conservative'' solution region 
is marked with thin dashed lines). 

\vspace{0.3cm}
%%%%%%%%%%%%%%%%%%%%%%%%%%%%%%%%
% CAPTION OF FIGURE : se_f50ab %
%%%%%%%%%%%%%%%%%%%%%%%%%%%%%%%%

\noindent {\bf Figures 10a -10b.} Iso - (D-N) asymmetry contour plots 
for the one year average ES {\it Night} (a) and {\it Core} (b)
asymmetries for the SNO detector  
in the case of the 
$\nu_e \rightarrow\nu_{s}$ transitions
of solar neutrinos for $\TeTh = 5.0$
and $\YeCore = 0.467$. 
The MSW $\nu_e \rightarrow\nu_{s}$ transition  
solution regions \cite{PK99} 
are also shown (the ``conservative'' solution 
region is marked with thin dashed lines). 

\vspace{0.3cm}
%%%%%%%%%%%%%%%%%%%%%%%%%%%%%%%%
% CAPTION OF FIGURE : sn1f__ab,sn1c__a_ %
%%%%%%%%%%%%%%%%%%%%%%%%%%%%%%%%

 \noindent {\bf Figures 11a - 11b.}
The same as in Fig. 10 
% Iso - (D-N) asymmetry contour plots 
for the one year average NC {\it Night} (a)
and {\it Core} (b) asymmetries
% for the SNO detector  
in the case of the 
$\nu_e \rightarrow\nu_{s}$ transitions
of solar neutrinos.
The solid (dashed) lines correspond 
to $\YeCore = 0.467~(0.500)$.

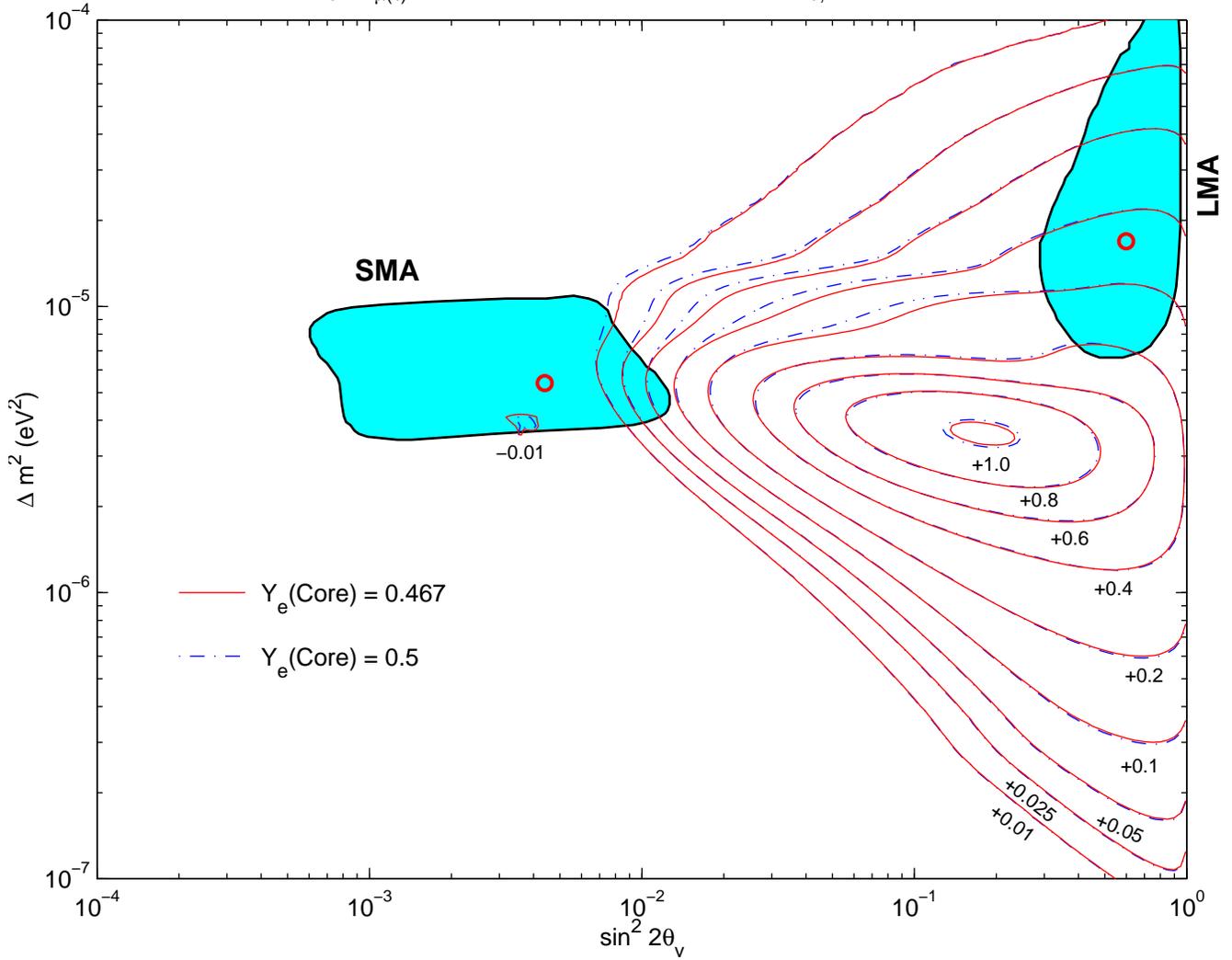
\end{document}